%% file: main.tex
\documentclass[10pt,twocolumn,letterpaper]{article}
\usepackage[pagenumbers]{cvpr}
\input{preamble}

\def\paperID{}
\def\confName{}
\def\confYear{}
\title{\ourtitle}
\author{
Hou In Ivan Tam$^{1}$ \quad Hou In Derek Pun$^{1}$ \quad Austin T. Wang$^{1}$ \quad Angel X. Chang$^{1,2}$ \quad Manolis Savva$^{1}$\\
$^{1}$Simon Fraser University \quad $^{2}$Alberta Machine Intelligence Institute (Amii)\\
\small\href{https://3dlg-hcvc.github.io/smc/}{3dlg-hcvc.github.io/smc}
}

\begin{document}

\input{fig/teaser}

\twocolumn[{
\maketitle
\figfirstpagefigure
}]

\input{sec/00-abstract}
\input{sec/01-intro}

\input{sec/02-related}
\input{sec/03-method}
\input{sec/04-experiments}
\input{sec/05-results}
\input{sec/06-conclusion}

\input{sec/ack}

{
\small
\bibliographystyle{ieeenat_fullname}
\bibliography{main}
}

\maketitlesupplementary

\appendix

\input{supp/supplement}

\end{document}

%% file: preamble.tex
\usepackage[dvipsnames]{xcolor}

\usepackage{multirow}
\usepackage{svg}
\usepackage{caption}
\usepackage[cachedir=minted-cache,frozencache]{minted} 
\definecolor{cvprblue}{rgb}{0.21,0.49,0.74}
\usepackage[pagebackref,breaklinks,colorlinks,citecolor=cvprblue]{hyperref}

\newcommand{\mypara}[1]{\noindent\textbf{#1}}
\newcommand{\prompt}[1]{\texttt{\footnotesize #1}}

\newcommand{\ourtitle}{SceneMotifCoder: Example-driven Visual Program Learning\\for Generating 3D Object Arrangements}

\newcommand{\ours}{SceneMotifCoder\xspace}
\newcommand{\oursshort}{SMC\xspace}
\newcommand{\oursintro}{SceneMotifCoder (SMC)\xspace}

\newcommand{\symnaiveprog}{\ensuremath{P_0}\xspace}
\newcommand{\symmotifprog}{\ensuremath{P_\text{motif}}\xspace}
\newcommand{\symmetaprog}{\ensuremath{P}\xspace}

\newcommand{\naiveprog}{na\"ive program\xspace}
\newcommand{\motifprog}{motif program\xspace}
\newcommand{\metaprog}{meta-program\xspace}
\newcommand{\naiveprogcap}{Na\"ive Program\xspace}
\newcommand{\motifprogcap}{Motif Program\xspace}
\newcommand{\metaprogcap}{Meta-program\xspace}

%% file: fig/teaser.tex
\newcommand{\figfirstpagefigure}{
\vspace{-2em}
\begin{center}
\captionsetup{type=figure}
\includegraphics[width=\textwidth]{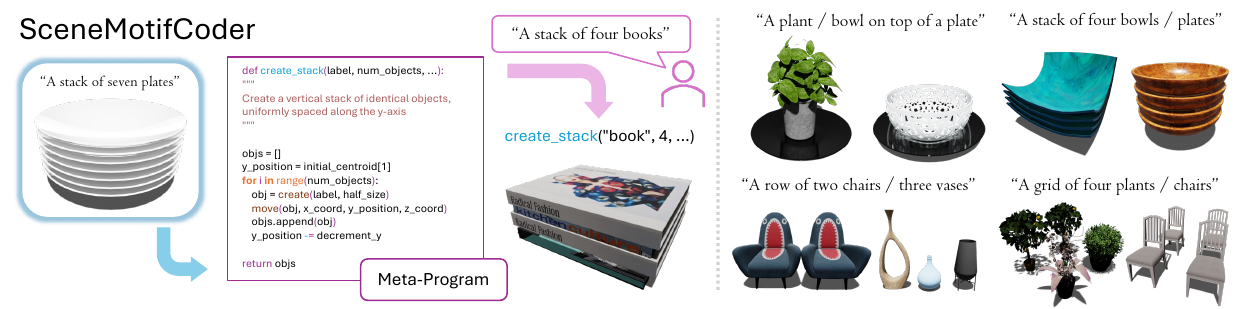}
\vspace{-2em}
\captionof{figure}{
We present \ours (\oursshort), an example-driven approach for generating 3D object arrangements.
\textbf{Left}: given an example arrangement and text description, \oursshort writes a meta-program with arguments that capture the underlying spatial patterns of the arrangement.
Then, to generate a new arrangement, \oursshort takes a text description as input, retrieves a meta-program, and writes a function call with appropriate arguments to execute the meta-program.
Through mesh retrieval and a geometry optimization step, the final arrangement is physically plausible and conforms to the text description.
\textbf{Right}: Example output 3D object arrangements for various input descriptions. \oursshort generates arrangements with programs learned from 1 to 3 examples.
}
\label{fig:teaser}
\end{center}
}

%% file: sec/00-abstract.tex
\begin{abstract}
Despite advances in text-to-3D generation methods, generation of multi-object arrangements remains challenging.
Current methods exhibit failures in generating physically plausible arrangements that respect the provided text description.
We present \oursintro, an example-driven framework for generating 3D object arrangements through visual program learning.
\oursshort leverages large language models (LLMs) and program synthesis to overcome these challenges by learning visual programs from example arrangements.
These programs are generalized into compact, editable meta-programs.
When combined with 3D object retrieval and geometry-aware optimization, they can be used to create object arrangements varying in arrangement structure and contained objects.
Our experiments show that \oursshort generates high-quality arrangements using meta-programs learned from few examples.
Evaluation results demonstrates that object arrangements generated by \oursshort better conform to user-specified text descriptions and are more physically plausible when compared with state-of-the-art text-to-3D generation and layout methods.
\end{abstract}

%% file: sec/01-intro.tex
\section{Introduction}

Digital 3D representations of indoor scenes are essential for many domains including interior design, game development, visual effects, virtual and augmented reality, and embodied AI simulations.
However, authoring these scenes is laborious and requires 3D modeling expertise.
Thus, indoor scene synthesis has been a research focus for more than a decade~\cite{merrell2011interactive,fisher2012example,chang2014learning}.
Recent progress enabled conditional scene generation given a room type~\cite{feng2024layoutgpt}, floor plan~\cite{paschalidou2021atiss}, or graph specifying objects and relations~\cite{wang2019planit}.

Despite this progress, existing scene synthesis methods focus on large furniture objects such as beds, cabinets, tables, and chairs and typically ignore smaller objects.
For example, generated scene layouts lack plates, cups or decorative objects on dining tables, making the scenes look unrealistically empty.
Such objects are ubiquitous in real life and are vital to making synthetic scenes more realistic.

Nevertheless, modeling these object arrangements is challenging.
Existing 3D scene datasets rarely capture small-scale arrangements, making standard generative model training strategies impractical.
Additionally, the highly compositional and geometrically tight nature of these arrangements complicates placement of objects.
For instance, plates in a stack of plates are in contact with plates above and below without gaps or intersections, requiring precise placement that is hard to achieve with existing layout generation methods.

In this paper, we present \emph{\ours}, an example-driven framework for 3D object arrangement generation based on visual program learning.
We show how the code generation capabilities of Large Language Models (LLMs) can be leveraged to create visual programs that capture abstractions of common 3D object arrangements, or \emph{motifs}, from few examples.
These motif visual programs then allow generation of new object arrangements exhibiting substantially different content (i.e. different objects) and structure (i.e. different numbers and positions of objects), from text descriptions provided by a user.

Our key insight is that object arrangements have underlying motifs that can be extracted from a few examples.
These motifs can be treated as templates for creating new arrangements.
To this end, programs are an ideal representation as the control flow compactly and interpretably captures the motif structure.
By generalizing motif programs into meta-programs, we further concretize common structures and parameters within a motif type and allow for easy instantiation of new arrangements.

We show that \oursintro requires only 1 to 3 examples to learn a meta-program.
By leveraging these re-usable meta-programs we can significantly streamline 3D content creation, especially for designers who routinely create layouts using a 3D object library.
Unlike black box 3D generative models, \oursshort's human-readable programs allow user modification to intuitively adjust generation results.
Furthermore, \oursshort's generation capability is easily extendable.
Given new examples, \oursshort learns new motifs without resource intensive and time consuming re-training.
Thus, the learned meta-program library can efficiently and progressively grow into a comprehensive knowledge base for a wide range of realistic object arrangements.

We evaluate \oursshort against state-of-the-art layout and text-to-3D generative models and show that our object arrangements better conform to input descriptions and exhibit higher physical plausibility, measured both algorithmically and through human perceptual studies.
In summary:
1) we propose the \oursshort framework, an example-driven approach for learning visual programs that represent 3D object arrangements;
2) we use the learned programs to generate complex and realistic arrangements for diverse objects and show how our framework admits for interpretable and easily editable arrangement generation using simple text descriptions; and
3) we systematically evaluate against state-of-the-art layout and text-to-3D generation methods and show our approach outperforms prior work in both alignment to text description and physical plausibility.

%% file: sec/02-related.tex
\section{Related Work}

We summarize prior work on various strategies for 3D scene synthesis: compositional 3D layout generation using object retrieval, layout generation leveraging LLMs, text-to-3D generation using differentiable rendering, and visual program synthesis methods.

\mypara{3D layout generation.}
There is much work on layout generation, ranging from rule-based~\cite{merrell2011interactive,yu2011make}, to data-driven~\cite{fisher2012example}, to neural network methods~\cite{wang2018deep}.
While early work~\cite{fisher2012example} focused on cluttered object arrangements, more recent work using deep learning~\cite{wang2018deep,wang2019planit,zhou2019scenegraphnet,wang2021sceneformer,paschalidou2021atiss,para2022cofs,tang2023diffuscene,lin2024instructscene} focuses on larger furniture-sized objects, typically ignoring smaller objects on tabletops or other surfaces.
One reason is that available 3D scene datasets mostly contain furniture and are sparsely populated with `detail objects'.
Our approach is tailored for generation of such detail object arrangements.

\mypara{Layout generation with LLMs.}
A recent trend uses LLMs for more open-vocabulary layout generation and generation conditioned on text input.
Early work relied on custom rules~\cite{coyne2001wordseye} or data-driven priors~\cite{chang2014learning,chang2017sceneseer} to determine spatial relations and retrieve objects with keyword search.
Recent work~\cite{wen2023anyhome,sun20233d,lin2024layoutprompter,feng2024layoutgpt,yang2024holodeck,huang2024blenderalchemy,hu2024scenecraft,zhou2024scenex,aguina2024open} leverages LLMs more flexibly 1) as a source of spatial priors, 2) to retrieve 3D assets based on vision-language embeddings, or 3) to generate modeling software API calls that generate 3D scenes.
In contrast, we combine LLMs with visual program synthesis to learn self-contained reusable programs that generate 3D object layouts given a small set of examples.
These programs capture arrangement \emph{motifs} that give the user more control and editability compared to relying on API calls directly generated by LLMs.

\mypara{Text-to-3D generation.}
Text-to-3D-shape generation models have advanced rapidly thanks to recent breakthroughs in text-to-image diffusion.
Various works demonstrated the use of image diffusion models to generate 3D shapes~\cite{poole2022dreamfusion,nichol2022point,wang2023score,yi2023gaussiandreamer,zhou2024dreamscene360}.
While some work attempted to generate scenes as single unstructured geometry~\cite{hollein2023text2room}, the output cannot be easily decomposed and manipulated as separate objects.
Thus, recent works generate multiple objects either from a prespecified layout~\cite{po2023compositional,lin2023componerf} or by tackling both the layout (often using an LLM to determine how objects should be arranged) and generation together~\cite{vilesov2023cg3d,gao2023graphdreamer,zhang2023scenewiz3d}.
Despite their success, these methods often generate incorrect numbers of objects, fail to respect spatial relationships, or suffer from the `Janus problem' (multiple implausible front sides).
Our work does not rely on a text-to-image diffusion model to obtain spatial relationship priors between objects.
Rather, given an object arrangement as an example we directly learn the spatial arrangement pattern as a visual program and use it to instantiate new object arrangements.

\mypara{Visual program synthesis.}
Visual programs are attractive for 3D shape representation as they are more interpretable and easily editable than low-level 3D representations.
In addition, they are easier to generate with LLMs due to the large volume of code in LLM training data.
Synthesis of visual programs has been applied to computer-aided design (CAD) sketches~\cite{ganin2021computer,para2021sketchgen,seff2021vitruvion} and CAD models~\cite{li2022free2cad,wu2021deepcad,xu2022skexgen}.
It has also been used to describe 3D object arrangements using program abstraction through interactive or in-context learning~\cite{wang2017naturalizing,liu2019learning}.
A recent line of work~\cite{jones2020shapeassembly,jones2021shapemod,jones2023shapecoder,ganeshan2023improving} applied visual programs to encode 3D shapes as primitives (e.g., a chair as a set of cuboids).
In particular, ShapeMOD~\cite{jones2021shapemod} and ShapeCoder~\cite{jones2023shapecoder} demonstrate programs can extract abstractions of high-level 3D object geometry patterns, facilitating downstream editing tasks.
Our work is similar to \citet{liu2019learning} in that we use visual programs to encode collections of objects.
However, we focus on the challenging task of learning dense 3D spatial arrangements given a text description and a few examples, rather than inferring programs from simple shapes (e.g., cuboids), and we generalize our inferred programs to create new arrangements.

%% file: sec/03-method.tex
\input{fig/overview}

\section{Method}

In our problem setting, a user provides examples of an object arrangement $A$ paired with text $T$ describing the arrangement.
The object arrangement consists of a set of labeled 3D assets with appropriate transformations.
Our method then takes these input examples consisting of a few $(T,A)$ pairs and produces a program $P$ that can generate new variations of arrangements (with different types of objects) for that motif type.
The learned program is stored in a library, to be retrieved at inference (i.e. generation) time, and used to generate new arrangements.
Our visual programs use a simple Domain Specific Language (DSL) based on Python (\cref{sec:dsl}).

Our approach, \oursintro, consists of two phases (see \cref{fig:overview}): the \emph{learning} phase (\cref{sec:learning}) where we extract visual programs from a few (1-3) examples, and the \emph{inference} phase (\cref{sec:inference}) where the program is retrieved and applied to a new text description.
In the \emph{learning} phase, we first represent the arrangement as a \emph{na\"ive program} that specifies the exact objects to be instantiated with their label, position, and orientation.
Then, we use an LLM to do a series of transformations that go from the list-based \emph{na\"ive program} to a more structured \emph{motif program} that specifies the given arrangement but uses higher-level constructs such as loops, to a final \emph{meta-program} (i.e. function with arguments) that we store in the meta-program library.
During \emph{inference}, we retrieve the appropriate meta-program and use an LLM to determine appropriate call arguments.
Based on the arrangement, we retrieve 3D objects and perform a geometry-aware optimization before outputting the final arrangement.
This strategy leverages the strong code generation capabilities of LLMs to produce the meta-programs, while still relying on the concrete mesh retrieval and geometric optimization to create the final arrangement.

\subsection{Domain Specific Language}
\label{sec:dsl}

We use Python as the basis for our DSL for its readability and high-level constructs.
In addition to built-in Python functions, the numpy\footnote{\url{https://numpy.org}} library and the following three constructs are available to describe motifs using programs:
\begin{enumerate}
    \item \texttt{create(label,size)} : Instantiate object of type \texttt{label} with dimensions \texttt{size} $\in\mathbb{R}^3$ at scene origin.
    \item \texttt{move(object,x,y,z)} : Position \texttt{object} at location $(x, y, z)\in\mathbb{R}^3$ in scene coordinates.
    \item \texttt{rotate(object,axis,angle)} : Orient \texttt{object} through rotation by \texttt{angle} around \texttt{axis} in object local coordinates.
\end{enumerate}
We avoid higher-level constructs such as relative position and relative rotation between objects, as we want to investigate whether an LLM can capture such relationships in the program structure and using arithmetic operations.

\input{fig/learning-details}

\subsection{Learning Phase}
\label{sec:learning}

\oursintro learns to write meta-programs to capture the underlying motifs of object arrangements that can be used to generate new arrangements in the learning phase.   
To do so, we generate and refine the program in three stages:
1) we first generate a \emph{na\"ive program} \symnaiveprog for each example arrangement based on the specific positioning and rotation observed in the arrangement;
2) we then prompt the LLM to make observations about patterns in the program to guide the LLM to rewrite the program to a more structured \emph{motif program} \symmotifprog; and
3) finally, given a small set of motif programs (of the same type), we ask the LLM to create a \emph{meta-program} \symmetaprog that captures commonalities across the motif programs, and extracts important axes of variation as parameters.
The top half of \cref{fig:overview} illustrates this phase.
For (2) and (3), as the LLM may make errors, we adopt an observe-generate-validate approach, in which we ask the LLM to make observations about the input, and after the LLM generates an initial program, we validate whether the generated program is executable and preserves the required functionality.
We iteratively prompt the LLM to refine the generated program until the validation is passed (typically within three iterations).
We also use chain-of-thought style prompting to encourage the LLM to reflect and reason about the program it is generating.
\Cref{fig:learning-details} shows a conceptual outline of these steps, which we describe in detail below.

\mypara{Na\"ive-program extraction.}
Given a text description and the corresponding object arrangement in individual meshes, \oursshort first extracts each object's label, bounding box centroid, and local coordinate axes from the input arrangement.
These attributes indicate the objects that exist in the arrangement, as well as their positions and local rotations.
Using these attributes, \oursshort writes a \textit{\naiveprog} \symnaiveprog for the arrangement by creating and posing each object one by one in a flat list of statements.
This program serves as the foundation for subsequent reasoning tasks.

\mypara{Motif-program generation.}
Using the \naiveprog \symnaiveprog, \oursshort then prompts an LLM to make four high-level observations about \symmotifprog.
These include:
1) the number of objects and their labels;
2) the relative displacements between objects;
3) whether there are any commonsense patterns (e.g., symmetry, repetition); and
4) whether there are any significant patterns in the x, y, and z coordinates.
These observations guide the LLM to look at useful patterns for extracting the underlying motif.
With these observations made, \oursshort then asks the LLM to divide the objects into groups, disentangling the overall motif into smaller spatial patterns that are easier to express as code.
Finally, \oursshort prompts the LLM to classify the arrangement into one of the motif types.
See the supplement for the LLM prompts.

The next step rewrites \symnaiveprog into a more structured \textit{\motifprog} \symmotifprog to better capture the motif's spatial patterns.
This more structured program makes it easier to reason about high-level patterns across different \symmotifprog for generating a final \metaprog that can handle variations.
We instruct the LLM to rewrite \symnaiveprog using arithmetic operations and programming constructs such as loops and conditions, while keeping object placements unchanged.

\textit{Validation.}
We then validate the rewritten program by executing it and checking against the original arrangement using the following criteria:
(1) no syntax error,
(2) no hard-coded list of object attributes,
(3) same number of objects per label,
(4) objects placed at same locations, and
(5) objects scaled and oriented the same.
These criteria ensure that: the program is executable, i.e. there are no Python interpreter errors (1), and the rewritten program is extracting the underlying motif instead of memorizing object placements (2).
We validate (2) by asking the LLM to judge the rewritten program.
Criteria (3) through (5) ensure the program does not deviate from the original arrangement, and are validated using simple logic checks and by computing volumetric Intersection-over-Union (IoU)s between the bounding boxes of the motif from the rewritten program and the original arrangement.
If any criterion fails, we provide feedback regarding what failed to the LLM (e.g., listing incorrectly placed object coordinates and corresponding correct coordinates), and try rewriting the program again.
These validations are crucial for the program to represent the input arrangement's motif precisely.
Once all criteria pass, \oursshort stores the rewritten program, a \textit{motif program}, in the program library under the classified motif type.

\mypara{Meta-program generation.}
To write a meta-program that captures the essence of the motif type, we need to reason about what \emph{changes} and what \emph{remains unchanged} across different object arrangement instances of the motif type.
To this end, \oursshort gathers all motif programs of the same type from the program library, and provides them as input to the LLM, asking the LLM to observe commonalities and differences between them.
If there is only one motif program in the library, we prompt the LLM to hypothesize different versions of the program and make observations based on them.
We further prompt the LLM to reason about why the programs are classified under the same motif type despite their differences.
The goal is to extract the fundamental spatial patterns of the motif type from the programs and ignore extraneous elements.
This then serves as guidance for writing the meta-program.

The last step in the learning phase is to write a generalized meta-program for the motif type with arguments that can be called to create a new motif of this type.
More specifically, \oursshort tasks the LLM to extract user-controllable parameters (e.g., number of objects) from the programs and include them as arguments in the meta-program function signature.
This ensures that the meta-program is flexible and can be used to generate new motifs given varying text descriptions.
In the prompt, we specifically ask the LLM to write docstrings and comments that explain the code and serve as guides for using the meta-program.

\textit{Validation.}
As in the program rewrite step, we validate the meta-program by checking it against all motif programs it generalizes.
We prompt the LLM to write one function call per program to execute the meta-program multiple times and check the following three criteria:
(1) no syntax error;
(2) same number of objects as in the corresponding motif program;
(3) relative directions between all pairs of objects match those in the motif program.
These validations ensure the meta-program captures a generalization of the spatial patterns of the motif type and that it can recreate the original motif programs.
Once the meta-program is validated, \oursshort stores it in the meta-program library to be used to generate new motifs.

\input{fig/inference-details}

\subsection{Inference Phase}
\label{sec:inference}

In the inference phase, given a text description as input, \oursshort generates a new object arrangement using previously learned meta-programs.
The steps involved are described below and outlined in \cref{fig:inference-details}.

\mypara{Retrieval.}
Using the text description, \oursshort first retrieves the meta-program of the same motif type in the library by classifying the description into one of the motif types through an LLM prompt that enumerates the motif types and provides the text of the description.

\mypara{Execution.}
To execute the meta-program, we feed it along with the description to an LLM to construct a function call with suitable arguments.
We check for syntax errors and execute the call to get a motif for the input text description.
To instantiate the motif into an object arrangement, \oursshort retrieves meshes from an object dataset using the label and the bounding box dimensions of each object.
We prompt the LLM for whether each of the objects can have multiple common orientations in real life scenarios.
If yes, we rotate each candidate mesh around its local axes to determine the best orientation for matching the object's bounding box.
Otherwise, we do not rotate.
For each object, we rank all available meshes based on the difference in dimensions between the mesh and the object's bounding box, and randomly pick one of the top five candidates as the mesh for the object.
Alternatively, we can retrieve meshes by finding the best match between an object description and renderings of the meshes using a pre-trained CLIP model (e.g., OpenCLIP~\cite{ilharco2021openclip}).
See the supplement for a discussion.

\mypara{Geometry-aware optimization.}
Since the retrieved objects may not match the specified dimensions, and the motif from executing the meta-program may contain inaccurate object placements, we optimize the final object placements with an iterative geometry-aware procedure.
We start with the first two objects in the arrangement and add a new object in each iteration, until all objects are processed.
In each iteration, we detect intersections between meshes and resolve them by moving the objects apart in the direction opposite to the contact points.
We then prompt the LLM using the text description to check whether objects in the described arrangement should be in contact with neighbouring objects.
If yes, we start an iterative ray-casting procedure to move the objects closer.
In each ray-cast iteration, we sample points on the surface of a newly added object facing the nearest neighbour, and shoot rays in the direction of the neighbouring object.
We use ray-mesh intersection tests to determine the intersection distances from the ray origins and move the object towards the ray direction by the minimum of the distances,
dampened by a factor that grows for each iteration.
Finally, once ray-cast iterations are finished, we cast rays from the object towards the ground plane to determine if it is supported by other objects.
We move the object to the ground if it is not supported.
This procedure repeats until all objects are optimized.
The resulting arrangement conforms to the text description and is physically plausible.

%% file: fig/overview.tex
\begin{figure*}
\includegraphics[width=\textwidth]{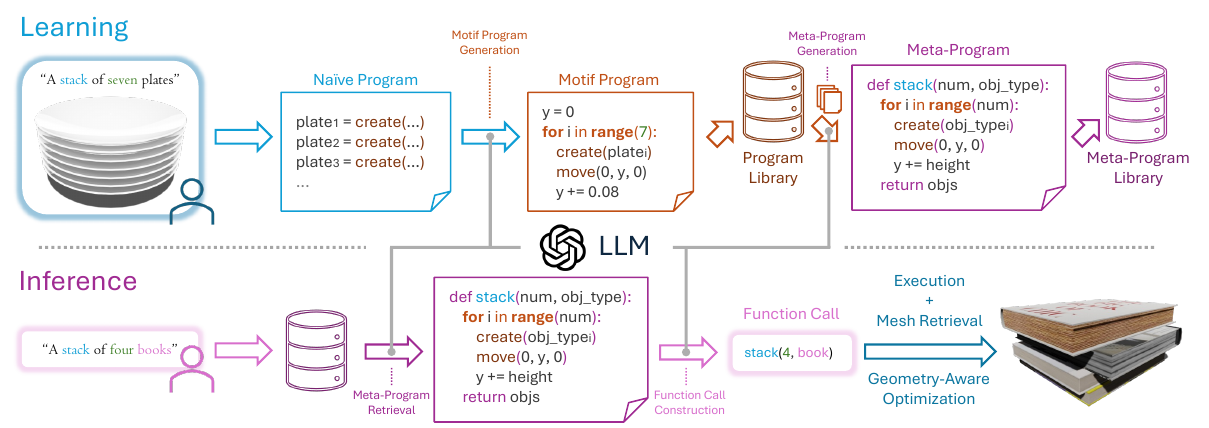}
\vspace{-1.75em}
    \caption{
    The \oursintro framework operates in two phases.
    \textbf{Top:} In the learning phase, given an object arrangement and its text description, \oursshort extracts a na\"ive program which is refined into a more structured motif program. The spatial patterns in the motif are then abstracted into a meta-program with reusable arguments to enable editing and generalization.
    \textbf{Bottom:} In the inference phase, given a text description as input, \oursshort selects a meta-program from the program library. Then, a new object arrangement is generated by writing a function call that executes the meta-program with appropriate arguments. Mesh retrieval and geometry-aware optimization then produces the final output which is a new object arrangement conforming to the input description.
    }
    \label{fig:overview}
\end{figure*}

%% file: fig/learning-details.tex
\begin{figure*}    
\vspace{-0.75em}
    \includegraphics[width=\linewidth]{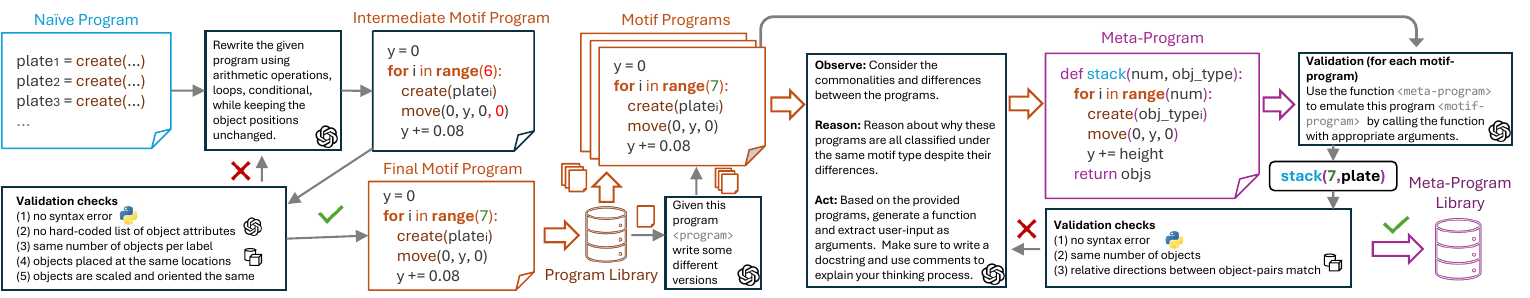}
\vspace{-1.75em}
    \caption{
    \ours learning phase overview.
    \textbf{Left}: A \naiveprog converted to a \motifprog and added to the program library. \textbf{Right}: The \motifprog is generalized through iterative refinement to a \metaprog and added to the meta-program library.
    }
    \label{fig:learning-details}
\vspace{-0.75em}
\end{figure*}

%% file: fig/inference-details.tex
\begin{figure}    
\includegraphics[width=\linewidth]{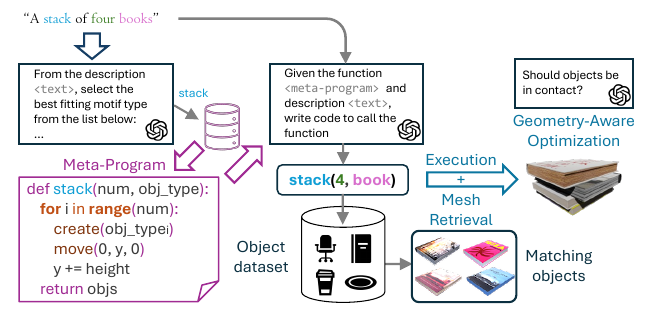}
\vspace{-2.0em}
    \caption{
    \ours inference phase overview.
    \textbf{From top left}: The input text is used to retrieve an appropriate meta-program from the library.
    Arguments are inferred to form a function call.
    The arrangement is then instantiated by executing the call, retrieving appropriate objects from a database, and optimizing the layout.
    }
    \label{fig:inference-details}
\vspace{-1.0em}
\end{figure}

%% file: sec/04-experiments.tex
\section{Experimental Setup}

We evaluate the \oursintro framework using 3D assets from the Habitat Synthetic Scenes Dataset (HSSD-200)~\cite{khanna2023hssd}.
HSSD consists of human-authored 3D scenes that closely mirror real-life indoor scenes, using a set of high quality, diverse 3D object assets.
The scenes contain a large variety of object arrangements which are ideal for learning real-world motifs.

To create a comprehensive evaluation of arrangement generation quality, we compile 181 everyday object arrangement descriptions belonging to 12 motif types: 
\texttt{stack}, \texttt{pile}, \texttt{row}, \texttt{grid}, \texttt{left\_of}, \texttt{in\_front}, \texttt{on\_top}, \texttt{surround}, \texttt{rectangular\_perimeter}, \texttt{wall\_column}, \texttt{wall\_row}, and \texttt{wall\_grid}.
We also include 21 descriptions for letter-shaped decorative arrangements resembling the letters \texttt{A}, \texttt{G}, \texttt{H}, \texttt{I}, \texttt{P}, \texttt{R}, and \texttt{S}.
In total, our test set includes 202 distinct input text descriptions.
These descriptions mention combinations of various everyday objects such as bowls, plates, cups, towels, books, and chairs.
There are a total of 50 unique object categories across all descriptions.
The everyday motifs are represented by between 5 and 27 descriptions, with a median of 16 descriptions per motif type, while the letter-shaped layouts each has 3 descriptions.

\oursshort learns a meta-program for each of the motif types given 1 to 3 example arrangements.
These examples are extracted directly from the HSSD scenes.
We manually annotate them with one text description each to serve as input for \oursshort's learning phase.
At generation time, we retrieve 3D objects from HSSD (not limited to the objects that were used in the examples).
We use GPT-4-Turbo (\textit{gpt-4-turbo-2024-04-09})~\cite{openai2023gpt} as the LLM for synthesizing visual programs.
The average cost for LLM calls per learned motif meta-program is approximately \$0.57.

\subsection{Baselines}

\oursshort generates tightly arranged spatial motifs of smaller objects, so we compare it with prior work that shares similar constraints.
In particular, we compare with two text-to-3D generative models: \textbf{MVDream}~\cite{shi2023mvdream} and \textbf{GraphDreamer}~\cite{gao2023graphdreamer}.
We choose MVDream as the representative approach for general text-to-3D generation as it excels at generating objects that are consistent across views, which is especially important in the multi-object setting.
GraphDreamer is a recent work that claims to excel at generating multi-object arrangements, making it the closest to our task.
It takes as input a scene graph generated by an LLM from text that describes the objects and their spatial relationships and optimizes an SDF for each object iteratively.
See supplement for baseline implementation details.

\subsection{Metrics}

There is not yet a clear consensus on how to evaluate 3D generative models.
GPTEval3D~\cite{wu2023gpteval3d} showed that GPT-4V is capable of evaluating 3D shapes along various criteria.
While GPTEval3D was not originally intended to evaluate layouts, we empirically observed that recent advances in LVLMs improved visual content perception and assessment of object placements.
Therefore, we report the text-asset alignment (\textbf{Align}) and 3D plausibility (\textbf{Plaus}) metrics from GPTEval3D.%
We also report the combination of these metrics (\textbf{Overall}).
The final score for \oursshort and the baselines are computed using 120 multi-view RGB images per output arrangement.
For these metric computations with GPTEval3D, we use GPT-4o instead of the default GPT-4-Vision-Preview and GPT-4-Turbo used for result generation, as we found GPT-4o to perform better at visual perception tasks which form the core of the evaluation.

In addition, we conduct manual verification to evaluate the generation results on three axes:
(1) correct number of objects (\textbf{\# Objs});
(2) same layout as in the text description (\textbf{Layout}); and
(3) the objects and their placements are physically plausible and appear the same as they would in real life (\textbf{Plaus}).
Given pairs of input text prompt and generated arrangement (each represented by two rendered views), an annotator evaluated each method's results under the above criteria by giving a binary judgment per criterion.

Lastly, we carry out a perceptual evaluation study to report human preference of the generated object arrangements.
The study is conducted using a two-alternative forced choice setup (A/B test).
We randomly pick 20 text prompts and create 60 questions.
Each question shows results from two of the three methods using the same input text description.
Participants are asked to judge which result is better along two axes:
(1) conformance to the text description --- i.e. how well the arrangement respects the objects and relations described in the text (\textbf{Align}); and
(2) realism --- i.e. physical plausibility of the arrangement (\textbf{Realism}).
We instruct participants to ignore texture and styling differences and focus solely on the above criteria.
The study was done with 31 participants not involved with this work.
We report the percentage of times each method was preferred for each description, compared against the other method.
See the supplement for details.

%% file: sec/05-results.tex
\section{Results}

\input{tab/quantitative}

\input{fig/qualitative}

\mypara{Comparison with prior work.}
\Cref{fig:qualitative} shows arrangements generated using \oursshort and the baselines.
\Cref{tab:quantitative} summarizes the quantitative evaluation.
To verify GPTEval3D's ability to evaluate object layouts, we computed the correlation of its pairwise comparisons against manual verification and user study results and found agreement of 88.5\% and 85.2\%, demonstrating high alignment with human judgment.
We also report the mean generation time.

The results show that \oursshort better conforms to the input text description in terms of the number of objects and the way they are arranged.
MVDream sometimes omits an entire object category, and disregards conditions on the number of objects.
It also sometimes generates extra objects (e.g., the floor in row 6).
GraphDreamer performs worse than MVDream as it almost always disregards the specified object numbers and layout.
Moreover, its outputs suffer from the Janus problem (e.g., generating bowls in irregular shapes).
Objects are often blended together with indistinguishable boundaries in between.
MVDream also suffers from the Janus problem but to a lesser extent.
These limitations hinder the practical usefulness of the arrangements in populating digital scenes.

In contrast, using only 1 to 3 examples \oursshort excels at generating results that respect the input text description.
Most generated arrangements have the correct number of objects, and the objects are arranged according to the description.
Even in cases where the example \oursshort learned from is significantly different from the inference input description in terms of object categories, object count, and dimensions, \oursshort successfully generalizes the underlying motif.
See the supplement for more qualitative examples.

\input{tab/ablation}

\mypara{Ablations.}
We compare against three ablated versions of \ours:
1) direct arrangement generation given descriptions of the DSL (w/o program); 
2) without step-by-step approach of making high-level observations to rewrite the program with feedback (w/o observations); and
3) replace meta-program with in-context learning using example arrangements as reference directly (w/o meta-program).

We also report results from LayoutPrompter~\cite{lin2024layoutprompter}, a recent work on LLM-based 2D graphics layout generation.
LayoutPrompter is similar to the ablated \oursshort without meta-program as it also uses in-context learning with a few examples.
We adapt its CSS-based formulation, adding 3D object pose attributes to generate 3D layouts, and use mesh retrieval to turn the layouts into object arrangements.

\Cref{tab:ablation} shows the quantitative results.
Without the generalization capabilities of the program learning module, the system has a hard time generating arrangements with structural variations in the motif (e.g., \texttt{surround}), and generated programs are often incorrect (less than 40\% of programs are correct, mostly easy motifs such as \texttt{left\_of}).
Without the step-by-step approach and feedback, generated meta-programs often fail to recognize patterns in the motif and exhibit na\"ive memorization, making them less adaptable to new input descriptions.
Without meta-programs, the system is free to change program parameters arbitrarily, making it more likely to write programs with erroneous parameters.
This is especially problematic when the target arrangement has more objects than the in-context examples, as there is no explicit constraint on where to put extra objects.
Similarly, LayoutPrompter struggles at placing objects realistically in 3D.
Floating objects and wrong rotations are the most common failure modes (see \cref{fig:qualitative}).
Even with LLMs trained on large corpora of 2D layouts, it is not straightforward to transfer such knowledge to 3D.

\input{fig/scene-populate}

\input{fig/editability}

\mypara{Applications.}
Each meta-program powerfully captures general motifs.
\Cref{fig:scene-populate} shows how meta-programs learned by \ours can be leveraged to efficiently and realistically populate indoor scenes.
\Cref{fig:editability} shows how meta-programs can be edited and adapted.
See supplement for more examples of meta-program generalization.

\mypara{Limitations.}
We focused on simple motifs and relied on a library of curated 3D objects (i.e. pre-aligned, sized, and categorized).
While the motifs we addressed are simple, our approach easily extends to more complex arrangements and using CLIP or text-to-3D generation can loosen the curated 3D objects requirement and enable more fine-grained retrieval.
See the supplement for examples of more complex arrangements and using CLIP for retrieval.
We can also use LVLMs to determine the size and front orientation of objects following prior work~\cite{yang2024holodeck}.
Another limitation is that we do not consider stylistic consistency between objects.

%% file: tab/quantitative.tex
\begin{table*}
\centering
\resizebox{\linewidth}{!}
{
\begin{tabular}{@{} l rrr rrr rr r @{}}
\toprule
& \multicolumn{3}{c}{Manual Verification} & \multicolumn{3}{c}{GPTEval3D~\cite{wu2023gpteval3d}} & \multicolumn{2}{c}{Perceptual Study} & \\ 
\cmidrule(l){2-4} \cmidrule(l){5-7} \cmidrule(l){8-9}
                                             & \# Objs $\uparrow$ & Layout $\uparrow$ & Plaus $\uparrow$  & Align $\uparrow$  & Plaus $\uparrow$  & Overall $\uparrow$ & Align $\uparrow$  & Realism $\uparrow$ & Time $\downarrow$ \\ \midrule
GraphDreamer~\cite{gao2023graphdreamer}      & 0.18               & 0.24              & 0.11              & 858.87            & 844.81            & 866.67             & 0.21              & 0.13               & 95 min            \\
MVDream~\cite{shi2023mvdream}                & 0.42               & 0.64              & 0.51              & 1083.57           & 1127.38           & 1089.83            & 0.44              & 0.54               & 43 min            \\
\ours (ours)                                 & \textbf{0.93}      & \textbf{0.90}     & \textbf{0.76}     & \textbf{1162.28}  & \textbf{1176.43}  & \textbf{1153.67}   & \textbf{0.85}     & \textbf{0.83}      & \textbf{2 min}    \\ \bottomrule
\end{tabular}
}
\vspace{-0.5em}
\caption{
Evaluation with manual verification of generated arrangement quality by a human expert annotator, algorithmic evaluation with GPTEval3D~\cite{wu2023gpteval3d}, and a perceptual user study.
We also report mean time to generate arrangement.
Our approach outperforms the baselines across all metrics, particularly in alignment to the text description, as measured by the \# Objs and Layout metrics.
}
\label{tab:quantitative}
\vspace{-1.0em}
\end{table*}

%% file: fig/qualitative.tex
\begin{figure}
\centering
\includegraphics[width=\linewidth]{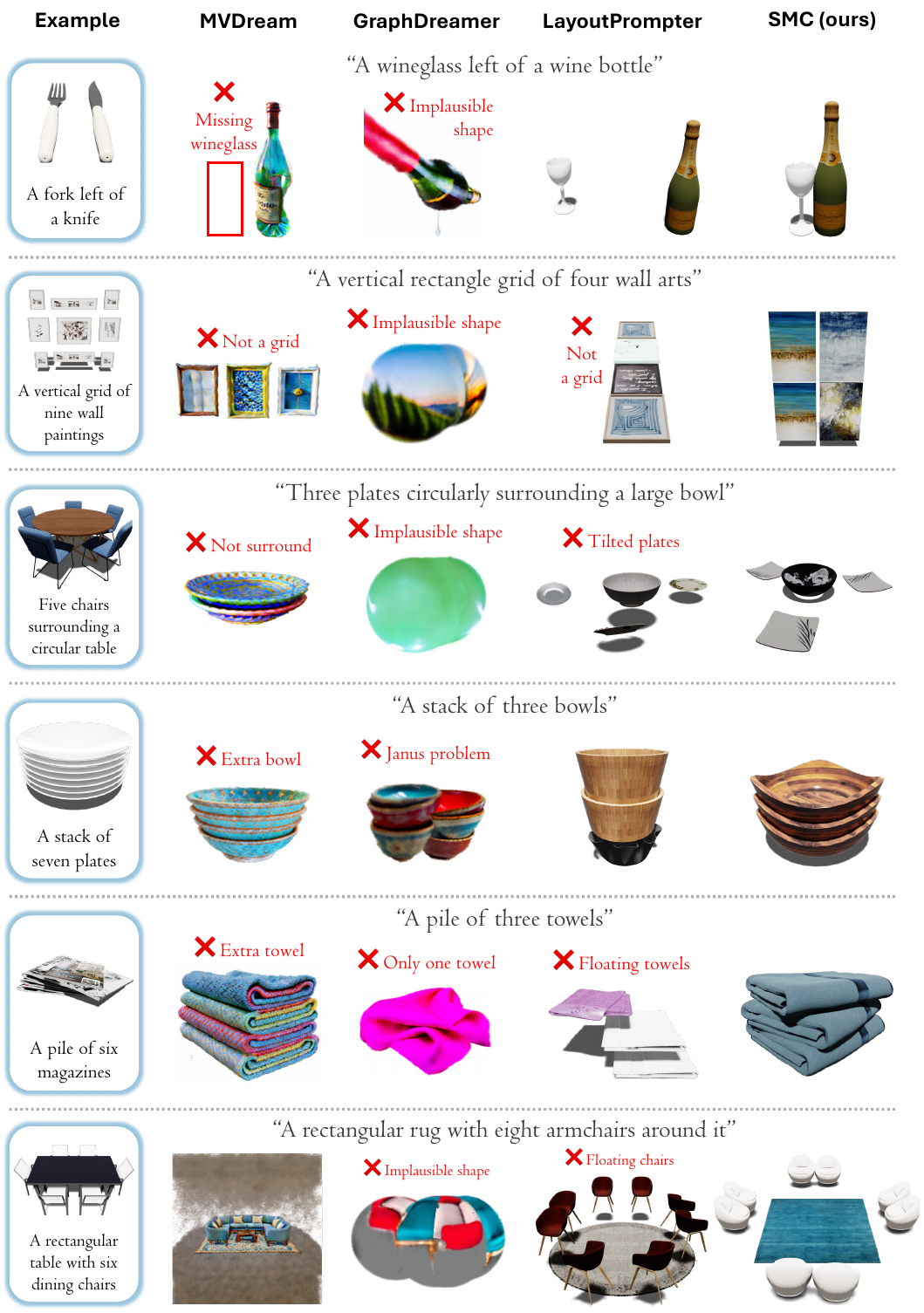}
\vspace{-1.75em}
    \caption{\textbf{Qualitative comparison of generated 3D object arrangements.}
    Each blue box on the left is the text and arrangement pair used for learning a meta-program capturing the arrangement's motif.
    The text above each row is the input at inference time.
    The arrangements generated by \ours better respect the input text in terms of number and type of objects present, and exhibit more plausible spatial configurations and object appearance.
    }
    \label{fig:qualitative}
\vspace{-0.75em}
\end{figure}

%% file: tab/ablation.tex
\begin{table}
\centering
\resizebox{0.9\linewidth}{!}
{
\begin{tabular}{@{}l ccc@{}}
\toprule
& \multicolumn{3}{c}{GPTEval3D} \\ 
\cmidrule(l){2-4}
& Align $\uparrow$ & Plaus $\uparrow$ & Overall $\uparrow$\\
\midrule
\ours (ours)                                & \textbf{1162.28}  & \textbf{1176.43} & \textbf{1153.67} \\
---~w/o meta-program                        & 1093.56           & 1119.31          & 1095.08          \\
---~w/o observations                        & 1080.74           & 1094.83          & 1076.70          \\
---~w/o program                             & 1036.18           & 1056.30          & 1035.42          \\
LayoutPrompter~\cite{lin2024layoutprompter} & 1067.13           & 1066.16          & 1057.53          \\
\bottomrule
\end{tabular}
}
\vspace{-0.5em}
\caption{
We show the impact of key modules in \ours.
LayoutPrompter and the \oursshort ablations generate arrangements with lower text-asset alignment, plausibility, and overall quality.
}
\label{tab:ablation}
\end{table}

%% file: fig/scene-populate.tex
\begin{figure}
\vspace{-0.75em}
\includegraphics[width=\linewidth]{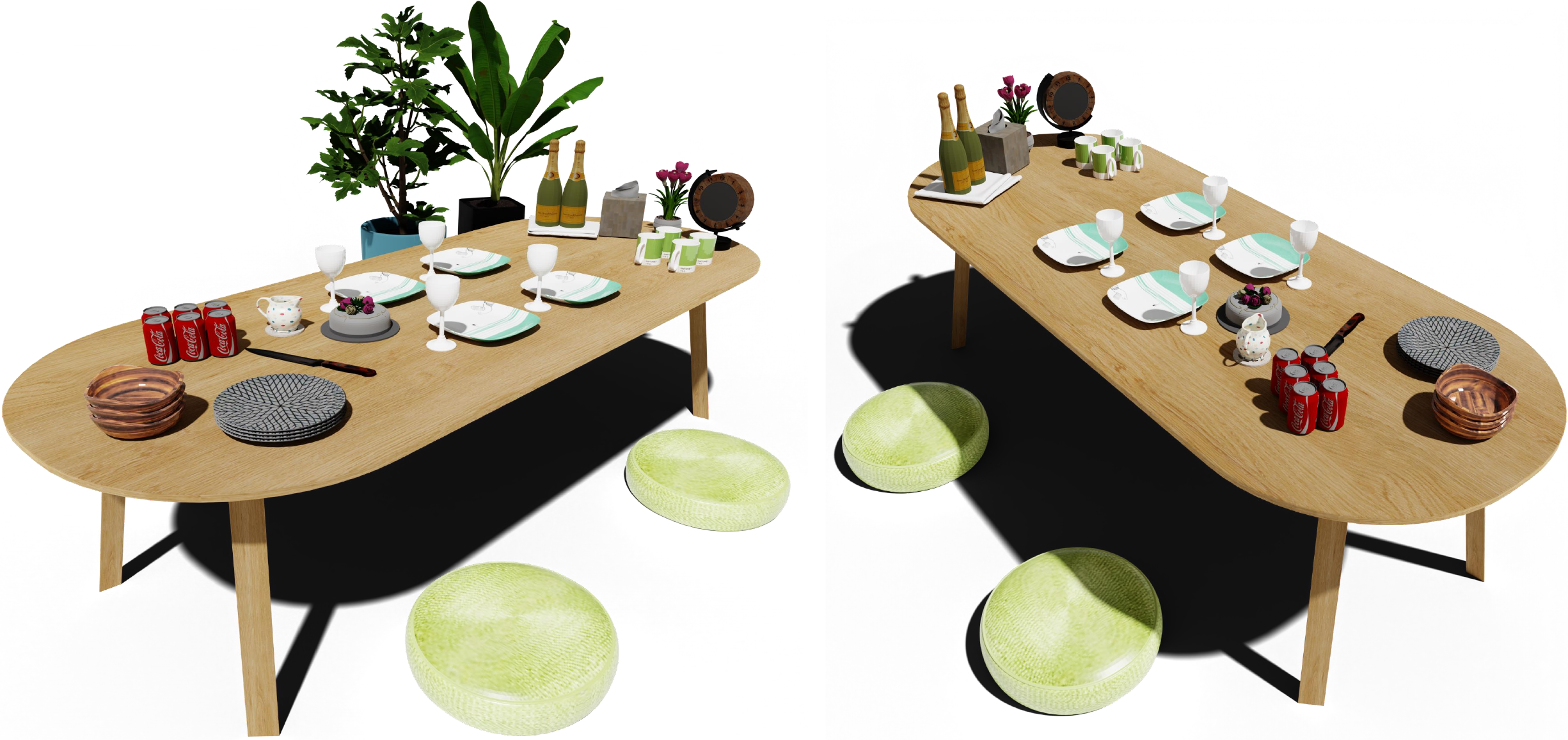}
\vspace{-1.75em}
\caption{
We populate the dining table scene shown with dense, realistic object arrangements by instantiating six learned motif types (\texttt{stack}, \texttt{row}, \texttt{grid}, \texttt{left\_of}, \texttt{in\_front}, and \texttt{on\_top}).
}
\label{fig:scene-populate}
\vspace{-0.4em}
\end{figure}

%% file: fig/editability.tex
\begin{figure}
\centering
\includegraphics[width=\linewidth]{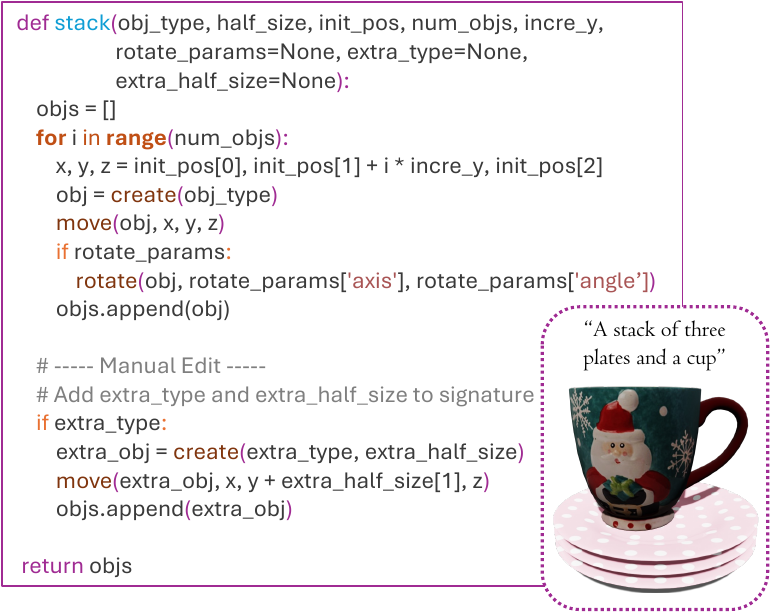}
\vspace{-1.5em}
\caption{
Editing a learned \texttt{stack} motif meta-program to add an object at the top of the stack (manually added code indicated in code comment).
The edited meta-program preserves the generality of the original while adapting the motif with a user modification.
}
\vspace{-1.0em}
\label{fig:editability}
\end{figure}

%% file: sec/06-conclusion.tex
\section{Conclusion}

We presented \oursintro, an example-driven framework for learning visual programs to generate 3D object arrangements.  While experts can code re-usable functions for specific arrangement, we showed that LLMs can automate this process and create reusable functions for common object arrangements.  Using LLMs can also be substantially cheaper and less time consuming.
We systematically evaluated our approach against state-of-the-art text-to-3D generation baselines, as well as ablations that show the value of our iterative visual program synthesis.
Our results show that arrangements produced by \oursshort have higher physical plausibility and respect the input text description more closely.
Using \oursshort, we enable users to easily generate and edit 3D object arrangements, and use them to populate indoor scenes.
We believe \ours's ease of use and efficiency for generating 3D object arrangements provides a scalable and flexible solution for populating 3D scenes with realistic object arrangements.

%% file: sec/ack.tex
\vspace{0.9em}
\mypara{Acknowledgements.}
This work was funded in part by a CIFAR AI Chair, a Canada Research Chair, NSERC Discovery Grant, and supported by \href{https://alliancecan.ca/}{Digital Research Alliance of Canada}.
We thank Qirui Wu, Jiayi Liu, and Han-Hung Lee for helpful discussions and feedback.

%% file: supp/supplement.tex
We provide additional implementation details for the baseline approaches (\cref{sec:supp-baseline-details}) and demonstration of how we can use OpenCLIP~\cite{ilharco2021openclip} for more fine-grained retrieval of matching objects (\cref{sec:supp-mesh-retrieval}).
We also provide evaluation details (\cref{sec:supp-eval-details}), additional examples of generated 3D arrangements (\cref{sec:supp-more-exmaples}), and generated programs (\cref{sec:supp-example-programs}).
Finally, we include the LLM prompts we used (\cref{sec:prompts}) and an example LLM session (\cref{sec:supp-session}).

\input{supp/details}

\input{supp/results}

\input{supp/programs}
\clearpage
\input{supp/prompts}
\input{supp/session}

%% file: supp/details.tex
\section{Implementation Details}
\label{sec:supp-impl-details}

\subsection{Baseline Details}
\label{sec:supp-baseline-details}

For MVDream, we use the default configuration without soft shading and lower the second stage resolution to 128 for computational efficiency.
Since GraphDreamer's public implementation only takes scene graphs as input and the prompts used in the paper are not available, we implement a text-to-scene-graph module using GPT-4 and follow the general strategy from the paper.
We also use the hyperparameter settings in GraphDreamer's demo scripts for all objects, except that we set center dispersion to 0.2 to allow the SDF for each object to be initialized at a reasonable distance from other objects.
Similar to MVDream, we lower GraphDreamer's 2nd stage resolution to 128 with batch size 1 to minimize VRAM usage.
Generating all 202 arrangements took a combined 152 GPU hours on three Nvidia L40S GPUs for MVDream, and 263 hours on four L40S and two A40 GPUs for GraphDreamer.
GraphDreamer often crashes due to out-of-memory errors for object arrangements that consist of more than three objects or contain larger objects.
For these arrangements, we re-run on two A100 GPUs with 80GB VRAM, which took another 181 GPU hours.
However, twelve input descriptions still crash due to out-of-memory errors.

\subsection{Mesh Retrieval Approach}
\label{sec:supp-mesh-retrieval}

Our work retrieves meshes to instantiate object arrangements based on object category and bounding box dimensions.
We tried incorporating OpenCLIP~\cite{ilharco2021openclip} to enable object appearance-based retrieval.
However, since our evaluation is focused on the generated object layouts, our test descriptions reflect that and do not specify appearance attributes such as colour.
As a result, the retrieved meshes do not differ significantly from the simpler approach when evaluated against the input text description.
Therefore, we opted for the simpler approach in our final system.
We note that in real life usage, users may have specific preferences to the appearances of the generated arrangements, making the feature-based retrieval module a worthwhile extension to our work.
See \Cref{fig:supp_retrieval} for a demonstrative example.

\input{fig/supp_retrieval}

\section{Evaluation Details}
\label{sec:supp-eval-details}

\subsection{Manual Verification Guidelines}
\label{sec:supp-manual-verification}

We provide the guidelines the annotator received below.

\begin{quote}
Looking at two views of each generation result, judge the generation quality based on the following three criteria. Give a binary judgment (Y/N) for each.
\begin{itemize}
    \item \# Objs - Correct number of objects as specified in the text description. Can you reasonably say there are $X$ instances of the object?
    \item Layout - Correct layout in terms of the relative positions between objects  as specified in the text description. Ignoring the appearances and labels of the objects, are the objects arranged correctly? If you were to classify the result into one of the motif types, will it fall into the same type as the description?
    \item Plaus - Physical Plausibility. Is the generation result physically plausible? Can the result exists in real life?
\end{itemize}
\end{quote}

\subsection{Perceptual Study Instructions}
\label{sec:supp-perceptual-study}

We provide the instructions the study participants received below.

\begin{quote}
This study asks you to judge the quality of 3D object arrangements in terms of 
\begin{itemize}
    \item Text Alignment - does the object arrangement fit the text description? E.g., for ``a stack of three books", an arrangement with two books stacked on top of each other does not fit the description.
    \item Realism - is the object arrangement realistic? You should disregard the fine-grained style and textured appearance of the objects. E.g., a chair that changes shape across view points and a teddy bear with five legs are unrealistic.
\end{itemize}
Judge the arrangements solely on these two aspects.
In each question you will see a text description followed by two object arrangements Left and Right, with a short video showing you different views of the arrangements.
Select the better arrangement according to the two aspects above.
Each question should take 5-10 sec.
There are 60 questions in total.
\end{quote}

%% file: fig/supp_retrieval.tex
\begin{figure}
    \includegraphics[width=\linewidth]{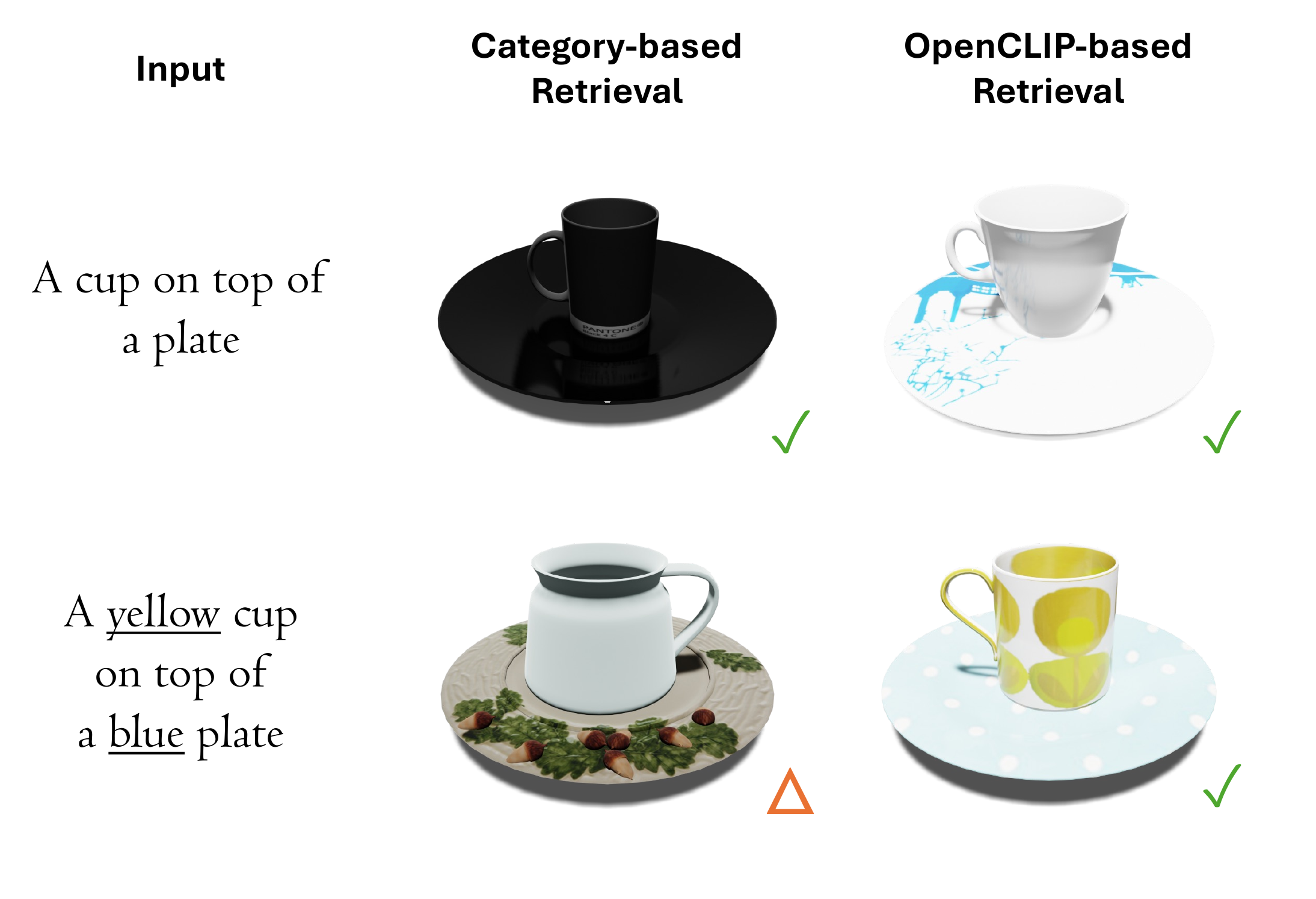}
    \caption{\textbf{Results using category-based and OpenCLIP-based retrieval.} First row: Input description does not specify object appearance. Both results are correct. This is the case for our test set descriptions. Second row: OpenCLIP-based retrieval is useful when the input description contains appearance attributes.}
    \label{fig:supp_retrieval}
\end{figure}

%% file: supp/results.tex
\section{Additional Qualitative Results}
\label{sec:supp-more-exmaples}

\input{fig/meta_example}

In \cref{fig:meta_example} we see how a meta-program can be called with varying text descriptions to produce quite distinct, yet realistic arrangements.
This is a concrete example of the re-usability and generality of our framework.

\Cref{fig:supp_qualitative} provides additional qualitative comparisons of generated object arrangements with prior work.
\Cref{fig:results} provides additional examples of arrangements generated using \ours.
The results showcase a variety of motif types, exhibiting different spatial arrangements.

In addition, \cref{fig:results_composition} shows how \ours's generated arrangements can be used as part of generating new compositional arrangements.
By treating previously generated arrangements as retrieval assets, \oursshort can invoke learned meta-programs to retrieve them and compose new arrangements with extra layers of complexity.
The results show that \oursshort can generate new arrangements that are significantly more complex than the original ones within a few iterations, demonstrating \oursshort's potential at generating complex and diverse arrangements.

\input{fig/supp_qualitative}

\input{fig/results}

\input{fig/results_composition}

%% file: fig/meta_example.tex
\begin{figure}
\includegraphics[width=\linewidth]{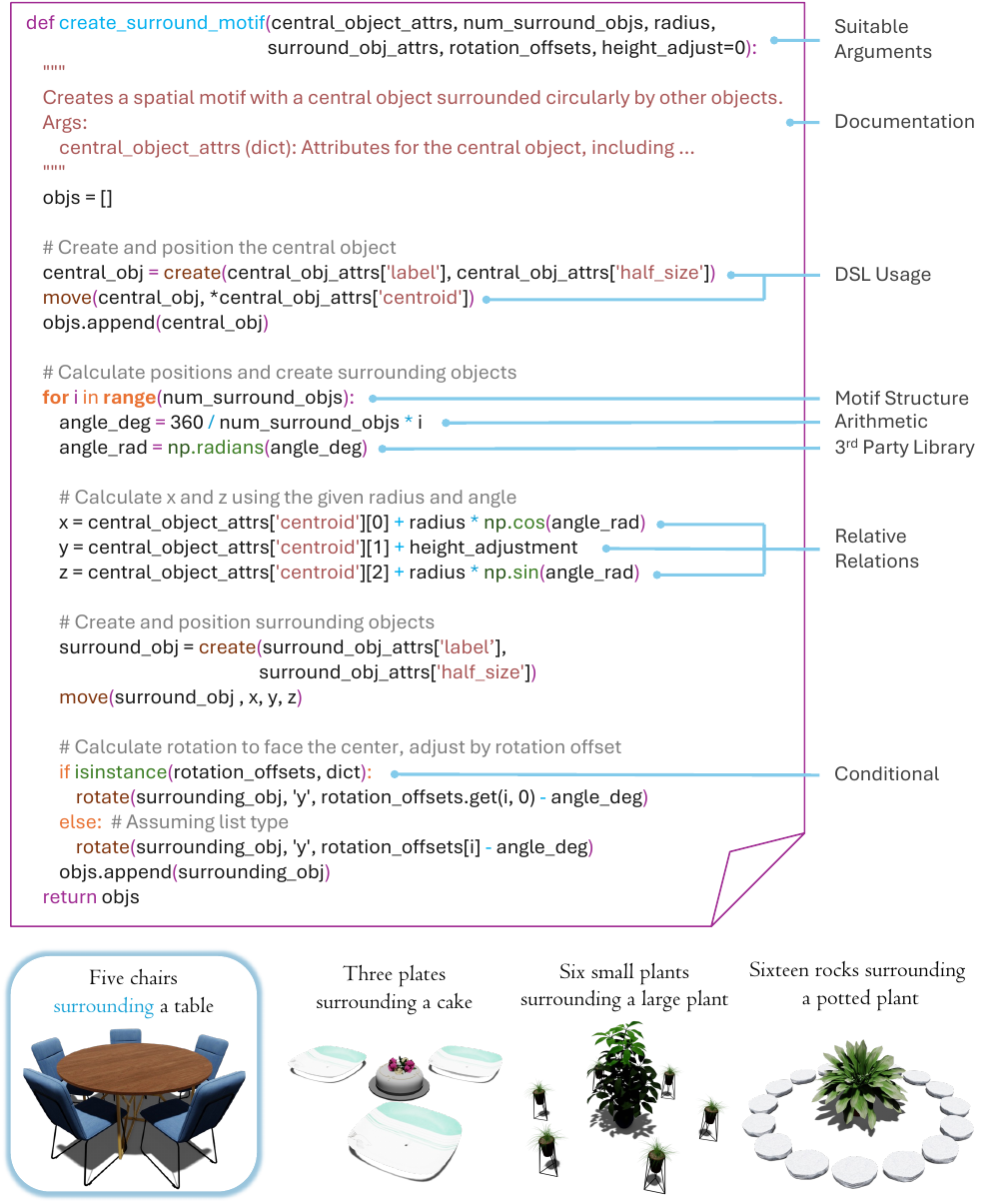}
\vspace{-2.0em}
    \caption{
    Meta-program for \texttt{surround} motif learned from the text and arrangement pair in blue box.
    The arrangements at the bottom are generated from the meta-program given the input text above each arrangement.
    Note the significant variations in object counts and types, all captured by human-interpretable and editable calls to the same meta-program.
    }
    \label{fig:meta_example}
\vspace{-0.5em}
\end{figure}

%% file: fig/supp_qualitative.tex
\begin{figure*}
\centering
\includegraphics[width=\textwidth]{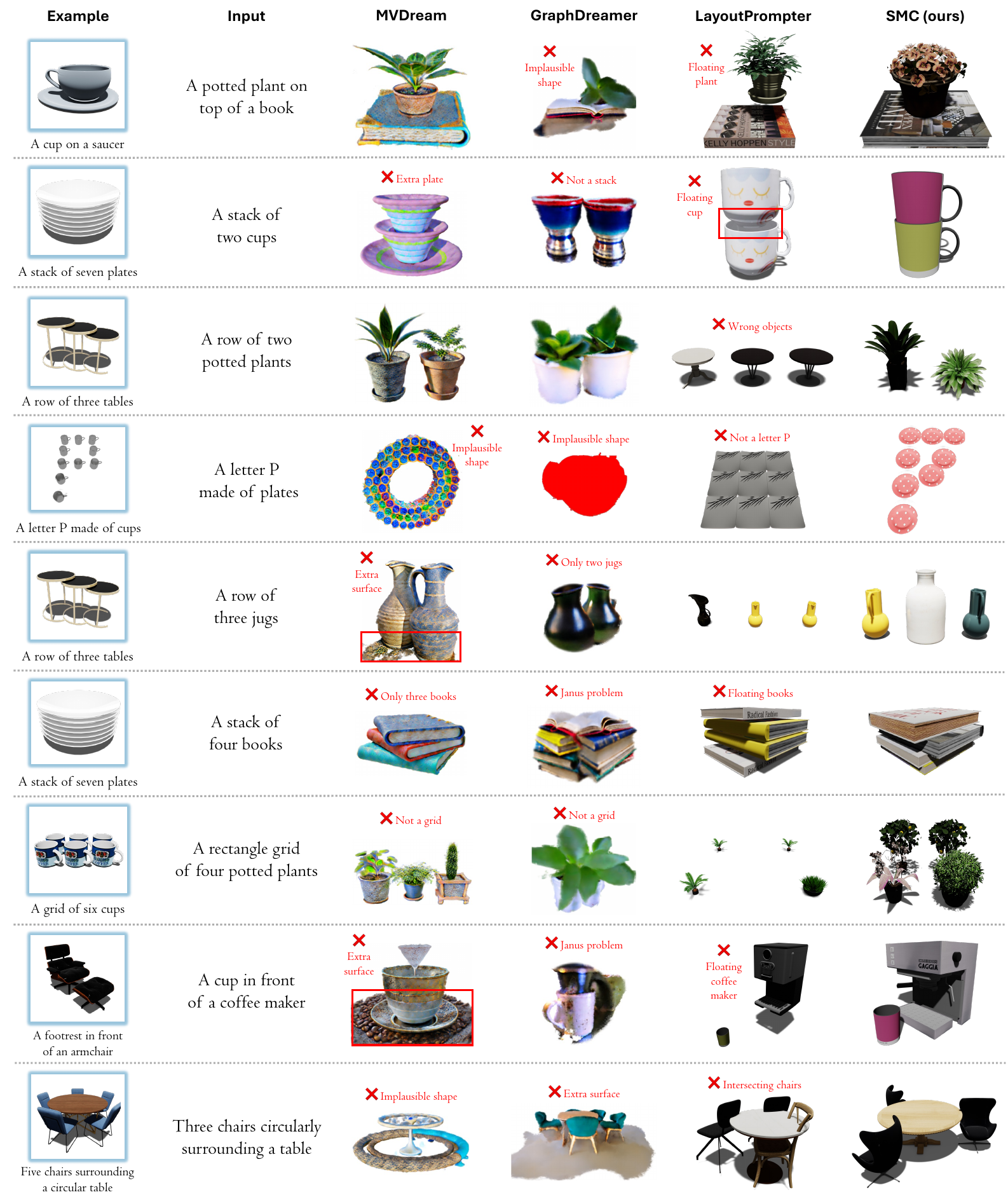}
    \caption{\textbf{Additional qualitative comparisons of generated 3D object arrangements.}
    Each blue box on the left is the text and arrangement pair used for learning a meta-program capturing the arrangement motif.
    The \ours arrangements better respect the input text in terms of number and type of objects present, and exhibit more plausible spatial configurations and object appearance.
    }
    \label{fig:supp_qualitative}
\end{figure*}

%% file: fig/results.tex
\begin{figure*}
\centering
\includegraphics[width=\textwidth]{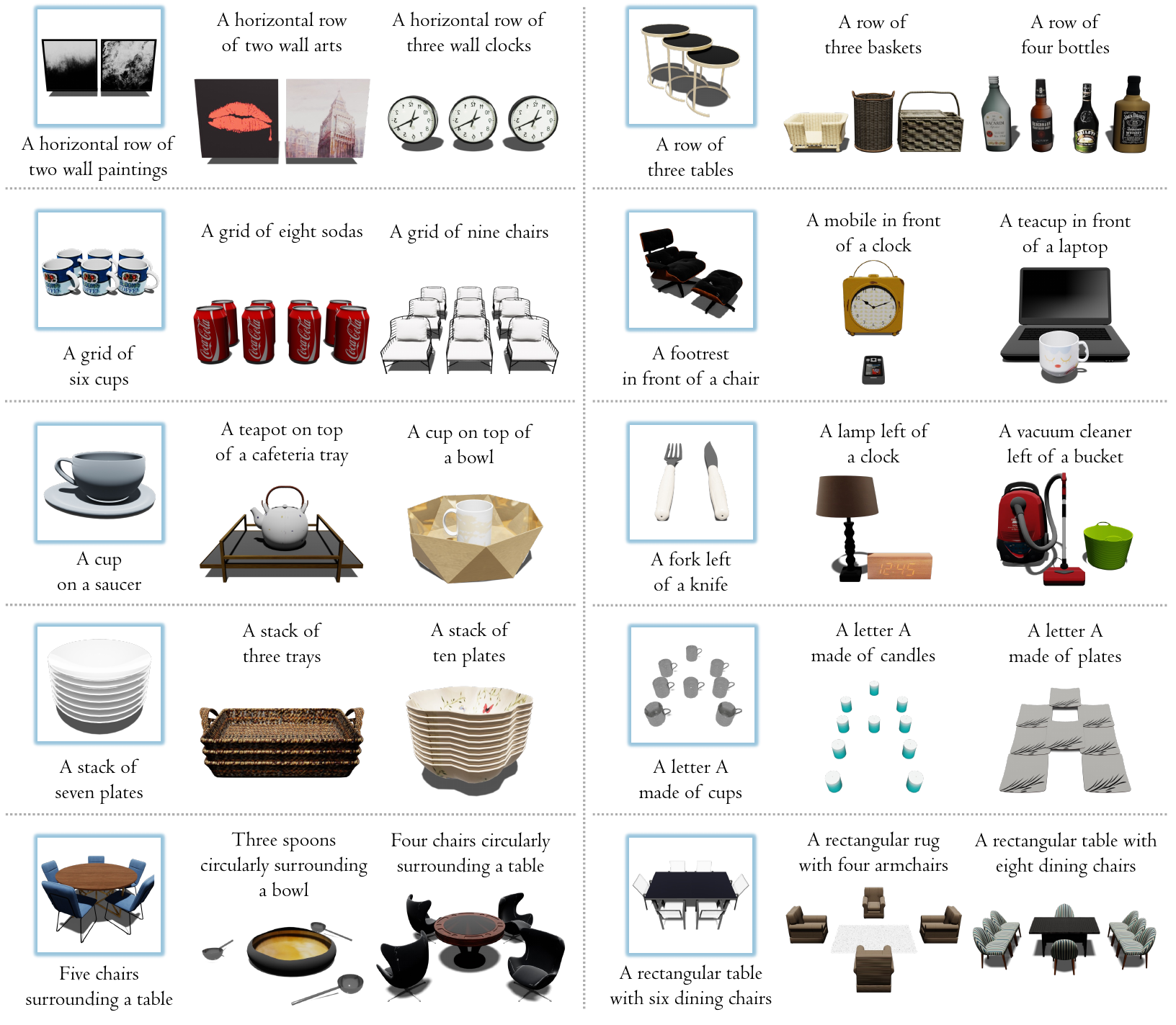}
    \caption{\textbf{Additional examples of generated object arrangements.}
    Our \ours framework generalizes the motifs exhibited by the text and arrangement example to produce diverse arrangements with differing object categories, counts, and spatial configurations.
    }
    \label{fig:results}
\end{figure*}

%% file: fig/results_composition.tex
\begin{figure*}
\centering
\includegraphics[width=\textwidth]{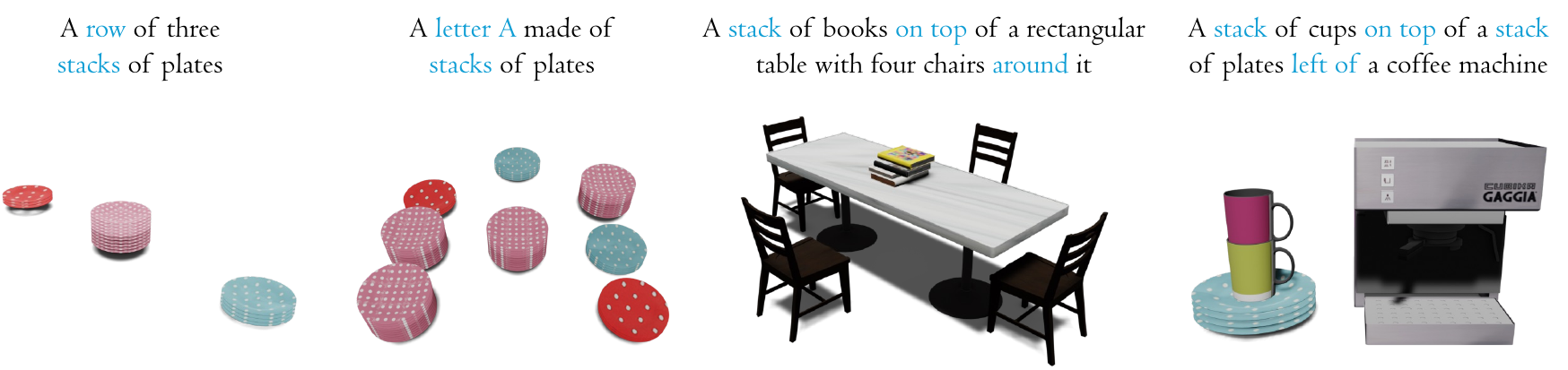}
\caption{
\textbf{Arrangements using compositional meta-programs.}
\ours's generated object arrangements can be used in generating new arrangements by treating them as retrieval assets.
The generated arrangements exhibit compounded layers of complexity within a few iterations.
Such compositions enable generation of more complex arrangements, opening the door to a wide range of possibilities.
}
\label{fig:results_composition}
\end{figure*}

%% file: supp/programs.tex
\section{Example Programs}
\label{sec:supp-example-programs}

In \Cref{tab:programs-pile,tab:programs-stack}, we provide example meta-programs generated by \ours, and compare them to two simpler prompting strategies:
\begin{itemize}
\item \emph{direct-from-description}: the LLM is prompted directly for a program from a description of a spatial motif, without information from example arrangements.
\item \emph{direct-from-motif-program}: the LLM is prompted to generate a meta-program given a set of motif programs, without any observation and reasoning steps.
\end{itemize}
The two \emph{direct-from-description} programs create na\"ive lists of repetitive statements with hard-coded values.
They cannot adapt to changes in the input description and cannot be reused to generate other arrangements.
While the \emph{direct-from-motif-program} programs are more flexible, they make strong assumptions on the object poses in their program structures.
These assumptions result in programs that lack generality, compared to the \ours meta-programs which are compact, human-readable, and admit many variations through setting of appropriate arguments in the function signature.
Analyzing the programs shows that 58\% of the \emph{direct-from-motif-program} programs, compared to only 5\% of meta-programs from \oursshort, have this issue.

These comparisons of concrete code examples help to demonstrate the value of the \ours visual program synthesis framework.

\begin{table*}[!ht]
\centering %
\begin{minipage}[t]{0.24\textwidth}
\textbf{direct-from-description}
\begin{minted}[breaklines, breakafter=d, fontsize=\scriptsize]{python}
# Initialize a list to hold all created objects
objs = []

# Create four seat cushions with labels and half sizes, appending them to objs list
objs.append(create("cushion1", [0.3, 0.1, 0.3]))
objs.append(create("cushion2", [0.3, 0.1, 0.3]))
objs.append(create("cushion3", [0.3, 0.1, 0.3]))
objs.append(create("cushion4", [0.3, 0.1, 0.3]))

# Set initial position for the first cushion
move(objs[0], 0, 0.1, 0)

# Stack the second cushion on top of the first
move(objs[1], 0, 0.3, 0)  # Additional 0.2 meters up from the first cushion's top surface

# Stack the third cushion on top of the second
move(objs[2], 0, 0.5, 0)  # Additional 0.2 meters up from the second cushion's top surface

# Stack the fourth cushion on top of the third
move(objs[3], 0, 0.7, 0)  # Additional 0.2 meters up from the third cushion's top surface
\end{minted}
\end{minipage}
\begin{minipage}[t]{0.35\textwidth}
\textbf{direct-from-motif-program}
\begin{minted}[breaklines, breakafter=d, fontsize=\scriptsize]{python}
def create_pile(label_prefix, num_objects, half_size, base_position, y_gap, rotation_angle):
    """
    Creates a spatial motif of type 'pile' with the given properties.

    Args:
    - label_prefix (str): prefix for the label of each object in the pile.
    - num_objects (int): number of objects to create in the pile.
    - half_size (list[float]): list of three floats representing half the size (x, y, z) of each object.
    - base_position (tuple[float]): starting (x, y, z) position for the base of the pile.
    - y_gap (float): additional gap between each object in the pile on the y-axis.
    - rotation_angle (float): angle to rotate alternating objects, to add naturalism.

    Returns:
    - list: list containing each created and positioned object.
    """
    objs = []
    base_x, base_y, base_z = base_position

    for i in range(num_objects):
        # Calculate the y offset for each subsequent object to simulate the pile
        new_y = base_y + i * (half_size[1] * 2 + y_gap)
        
        # Create an object with a unique label based on index
        obj = create(f"{label_prefix}_{i+1}", half_size)
        
        # Move the object to its designated position
        move(obj, base_x, new_y, base_z)
        
        # Alternate rotation for a more natural-looking pile
        if i %
            rotate(obj, 'z', rotation_angle)
        else:
            rotate(obj, 'z', -rotation_angle)
        
        # Append the positioned object to the list
        objs.append(obj)

    return objs
\end{minted}

\end{minipage}
\begin{minipage}[t]{0.4\textwidth}
\textbf{Meta-program from \ours}
\begin{minted}[breaklines, breakafter=d, fontsize=\scriptsize]{python}
import random

def create_pile_motif(object_type, num_objects, base_position, size_variations, position_offsets, rotations=[], seed=None):
    """
    Creates a spatial motif of type 'pile' with configurable positions, sizes, and rotations.
    
    Args:
    - object_type (str or list[str]): The type label(s) for the objects in the pile.
    - num_objects (int): The number of objects to include in the pile.
    - base_position (list of float): The starting position [x, y, z] for the first object in the pile.
    - size_variations (list of list of float): The dimensions [width, height, depth] for each object.
    - position_offsets (list of list of float): Position offsets [x, y, z] relative to the base for each object.
    - rotations (list of tuples): Rotations to apply, each as (axis, angle, index).
    - seed (int, optional): Seed for the random number generator for reproducibility.
    
    Returns:
    - list: A list of objects representing the created pile.
    """
    if seed is not None:
        random.seed(seed)

    objs = []

    # Iterate over each object to be created
    for i in range(num_objects):
        # Determine the object type if a list is provided
        obj_type = object_type[i] if isinstance(object_type, list) else object_type
        
        # Calculate size and position for the current object
        obj_half_size = size_variations[i]
        pos_offset = position_offsets[i]
        position = [base_position[0] + pos_offset[0], base_position[1] + pos_offset[1], base_position[2] + pos_offset[2]]
        
        # Create the object
        obj = create(obj_type, obj_half_size)
        move(obj, *position)
        
        # Apply rotations if specified for this object
        for rotation in rotations:
            if rotation[2] == i:  # Check if the rotation is meant for the current object
                rotate(obj, rotation[0], rotation[1])
        
        # Append the object to the list
        objs.append(obj)

    return objs
\end{minted}
\end{minipage}
\caption{
Different programs for generating ``a pile of four seat cushions''.
The \textit{direct-from-description} program is not reusable for generating new arrangements with different number of objects and types.
The \textit{direct-from-motif-program} program is reusable, but it makes a strong assumption on how to orient the objects (only alternating rotations around the z axis).
In contrast, \ours's program is less restrictive and takes into account for more potential object poses, which is especially important for the \texttt{pile} motif type.
}
\label{tab:programs-pile}
\end{table*}

\begin{table*}[!ht]
\centering %
\begin{minipage}[t]{0.24\textwidth}
\textbf{direct-from-description}
\begin{minted}[breaklines, breakafter=d, fontsize=\tiny]{python}
# List to keep track of all created objects
objs = []

# Create the table
table = create("table", [1.5, 0.1, 1])
objs.append(table)

# Move the table to the center of the scene (assume it is at coordinate (0, 0, 0))
move(table, 0, 0, 0)

# Create the chairs and place them around the table
# Chair size is arbitrarily chosen to be smaller than half the table size

# Create chairs
chairs = [create(f"chair_{i}", [0.3, 0.3, 0.3]) for i in range(8)]
objs.extend(chairs)

# Positions for chairs around the table
positions = [
    (1.8, 0.3, 0),    # right
    (-1.8, 0.3, 0),   # left
    (0, 0.3, 1.3),    # bottom
    (0, 0.3, -1.3),   # top
    (1.3, 0.3, 1.3),  # bottom-right
    (1.3, 0.3, -1.3), # top-right
    (-1.3, 0.3, 1.3), # bottom-left
    (-1.3, 0.3, -1.3) # top-left
]

# Move and rotate chairs
for i, (x, y, z) in enumerate(positions):
    move(chairs[i], x, y, z)
    if i in [0, 1]:  # right or left chair
        rotate(chairs[i], 'y', 90)
    elif i in [2, 3]:  # bottom or top chair
        rotate(chairs[i], 'y', 0)
    elif i in [4, 7]:  # corner chairs (right side)
        rotate(chairs[i], 'y', 45)
    elif i in [5, 6]:  # corner chairs (left side)
        rotate(chairs[i], 'y', -45)
\end{minted}
\end{minipage}
\begin{minipage}[t]{0.3\textwidth}
\textbf{direct-from-motif-program}
\begin{minted}[breaklines, breakafter=d, fontsize=\tiny]{python}
def create_rectangular_perimeter_motif( center_obj_label, center_obj_half_size, perimeter_obj_label, perimeter_obj_half_size, num_perimeter_objs, center_position):
    """
    Create a spatial motif where a central object is surrounded by a number of perimeter objects 
    in a rectangular perimeter pattern.

    Args:
    center_obj_label (str): The label for the central object.
    center_obj_half_size (list[float]): The half size of the central object [hx, hy, hz].
    perimeter_obj_label (str): The label for the perimeter objects.
    perimeter_obj_half_size (list[float]): The half size of the perimeter objects [hx, hy, hz].
    num_perimeter_objs (int): The number of perimeter objects.
    center_position (list[float]): The position [x, y, z] of the central object.

    Returns:
    list: List of all created objects (central and perimeter).
    """
    objs = []

    # Create the central object
    center_obj = create(center_obj_label, center_obj_half_size)
    move(center_obj, center_position[0], center_position[1], center_position[2])
    objs.append(center_obj)

    # Calculate positions for perimeter objects
    center_x, center_y, center_z = center_position
    cx, cy, cz = center_obj_half_size
    px, py, pz = perimeter_obj_half_size
    perimeter_positions = []

    # Calculate the rectangular perimeter
    spacing_x = cx + px + 0.1  # Additional spacing
    spacing_z = cz + pz + 0.1  # Additional spacing

    for i in range(num_perimeter_objs):
        if i %
            sign = -1 if (i // 2) %
            x = center_x + sign * (spacing_x + px)
            z = center_z + (i // 4) * spacing_z
        else:  # Front and back sides
            sign = -1 if ((i - 1) // 2) %
            x = center_x + ((i - 1) // 4) * spacing_x
            z = center_z + sign * (spacing_z + pz)

        perimeter_positions.append((x, center_y, z))

    # Create and position the perimeter objects
    for pos in perimeter_positions:
        perimeter_obj = create(perimeter_obj_label, perimeter_obj_half_size)
        move(perimeter_obj, pos[0], pos[1], pos[2])
        objs.append(perimeter_obj)

    return objs
\end{minted}

\end{minipage}
\begin{minipage}[t]{0.45\textwidth}
\textbf{Meta-program from \ours}
\begin{minted}[breaklines, breakafter=d, fontsize=\tiny]{python}
def create_rectangular_perimeter_motif(
    central_label, central_half_size, central_position,
    surrounding_label, surrounding_half_size,
    count_long_sides, count_short_sides,
    long_side_offset, short_side_offset,
    rotation_long_sides, rotation_short_sides,
    y_offset):
    """
    Create a spatial motif of type 'rectangular_perimeter' with the specified attributes.

    Args:
    - central_label (str): Label/type of the central object (e.g., 'table').
    - central_half_size (list of float): Half-size dimensions of the central object [x, y, z].
    - central_position (tuple of float): Position of the central object (x, y, z).
    - surrounding_label (str): Label/type of the surrounding objects (e.g., 'chair').
    - surrounding_half_size (list of float): Half-size dimensions of the surrounding objects [x, y, z].
    - count_long_sides (int): Number of surrounding objects on the longer sides (x-axis) of the central object.
    - count_short_sides (int): Number of surrounding objects on the shorter ends (z-axis) of the central object.
    - long_side_offset (tuple of float): Positional offsets (x, z) for objects on the longer sides.
    - short_side_offset (tuple of float): Positional offsets (x, z) for objects on the shorter ends.
    - rotation_long_sides (float): Rotation angle for objects on the longer sides.
    - rotation_short_sides (float): Rotation angle for objects on the shorter ends.
    - y_offset (float): Vertical offset to place all surrounding objects at a specified height.

    Returns:
    - list: List of all created objects.

    Example:
    # Example call to recreate the example program 1
    objs = create_rectangular_perimeter_motif(
        'table', [1.00012, 0.38418, 0.49848], (0.0, 0.0, 0.0), 'chair', [0.28401, 0.48686, 0.32049],
        2, 4, (1.01483, 0.01433), (0.41141, 0.62215), -90.0, 180.0, 0.06209
    )
    """
    objs = []

    # Create and place the central object
    central_obj = create(central_label, central_half_size)
    move(central_obj, *central_position)
    objs.append(central_obj)

    # Create and place objects on the longer sides
    for i in range(count_long_sides // 2):
        for sign in [-1, 1]:  # Use -1 and 1 to cover both sides
            surrounding_obj = create(surrounding_label, surrounding_half_size)
            # Compute x and z positions based on offsets and symmetry
            x_pos = central_position[0] + sign * long_side_offset[0]
            z_pos = central_position[2] + (i - count_long_sides // 4) * long_side_offset[1]
            move(surrounding_obj, x_pos, y_offset, z_pos)
            rotate(surrounding_obj, 'y', sign * rotation_long_sides)
            objs.append(surrounding_obj)

    # Create and place objects on the shorter ends
    for i in range(count_short_sides // 2):
        for sign in [-1, 1]:  # Use -1 and 1 to cover both sides
            surrounding_obj = create(surrounding_label, surrounding_half_size)
            # Compute x and z positions based on offsets and symmetry
            x_pos = central_position[0] + (i - count_short_sides // 4) * short_side_offset[0]
            z_pos = central_position[2] + sign * short_side_offset[1]
            move(surrounding_obj, x_pos, y_offset, z_pos)
            if sign == 1:
                rotate(surrounding_obj, 'y', rotation_short_sides)
            objs.append(surrounding_obj)

    return objs
\end{minted}
\end{minipage}
\caption{
Different programs for generating ``a rectangular table with eight chairs around it''.
The \textit{direct-from-description} program is not reusable for generating new arrangements with different number of objects and types.
The \textit{direct-from-motif-program} program is reusable, but it assumes equal spacing between objects (hard-coded as 0.1) and evenly-distributed objects around the center object, limiting what it can generate.
In contrast, \ours's program is less restrictive and allows for a wide range of customization.
}
\label{tab:programs-stack}
\end{table*}

\clearpage

%% file: supp/prompts.tex
\section{LLM Prompts}
\label{sec:prompts}
We provide the LLM prompts used for various phases in our \ours framework.  We provide general prompts in \Cref{sec:prompts-general}, prompts for generating motif programs from na\"ive programs in \Cref{sec:prompts-motif-program}, prompts for generating meta-programs from sets of motif-programs in \Cref{sec:prompts-meta-program}, inference prompts in \Cref{sec:prompts-inference}, and prompts for our ablation study in \Cref{sec:prompts-ablation}.  For the motif-program and meta-program generation, we have prompts that ask the LLM to make high-level observations about the input, to generate the motif or meta-program, validate the generated program, and feedback prompts to ask the LLM to iterate if the validation fails.  For the meta-program generation, we also have reasoning prompts that ask the LLM to reason about the arguments and structure of the function to be generated.

\subsection{General Prompts}
\label{sec:prompts-general}

We have two general prompts:
\begin{itemize} 
\item System introduction prompt specifying the DSL used for our visual programs.
\item Classification prompt for identifying the scene motif type given a description.  This \prompt{classify} prompt is used both during inference and as part of the motif-program generation to determine the motif type that the generated program should be stored under.
\end{itemize}

\begin{minted}[breaklines, breakafter=d, fontsize=\scriptsize]{yaml}
# Define prompts used when prompting LLM.
system: >-
  You are a Python programmer.
  You have access to all Python built-in functions and libraries that are available in Python 3.10.
  To use a function or a library, you need to import it first.
  However, you are not allowed to use any third-party libraries except numpy, which is available as np.
  Everything you write will be content inside a function, unless otherwise specified.
  Do not write the function signature or any other code outside the function, unless you are explicitly asked to do so.
  If the prompt asks you to respond with code only, you should not include any example usage nor text description in your response.
  Write comments to describe your thinking process.
  Your responses should strictly follow the requirements specified in the prompts.
  You will work with a domain-specific language (DSL) for describing spatial motif of objects.
  You will be asked to perform some tasks related to spatial motif programs written using this DSL.
  All functions in the DSL are valid Python functions.
  Here are the functions that exists:
  1. create(label: str, half_size: list[float]) -> Obj: Create a new object with the given `label` and `half_size`. The `half_size` is a list of three floats representing the half of the size of the object in the x, y, and z dimensions.
  2. move(obj: Obj, x: float, y: float, z: float) -> None: Move the object `obj` to the world coordinate (x, y, z). The object's position is the centroid of the object's bounding box.
  3. rotate(obj: Obj, axis: str, angle: float) -> None: Rotate the object `obj` around its local axis (x, y, or z) by the given angle.
  Units are in meters for x, y, z dimensions and degrees for angle.
  The world is in a right-handed coordinate system, that is, when looking from the front, the x-axis is to the right, the y-axis is up, and the z-axis is towards the viewer.
  When looking from above, the origin (0, 0, 0) is at the top-left corner of the screen, and the x-axis increases to the right, the z-axis increases towards the bottom of the screen.
# -------------------------------------------
classify: >-
  Based on your observations, classify the spatial motif of objects "<DESCRIPTION>" into one of the following motif types:
  1. stack - multiple objects of the same type are placed orderly on top of each other
  2. pile - objects are placed on top of each other but not in an orderly manner (when in doubt between stack and pile, choose pile if the description contains the word "pile" explicitly)
  3. row - objects are placed next to each other in a row
  4. grid - objects are placed orderly in a grid, like a chessboard
  5. left_of - one object placed to the left of another object
  6. in_front_of - one object placed in front of another object
  7. on_top - one object placed on top of another object
  8. surround - objects are placed around a central object in a circular manner
  9. wall_vertical_column - objects are placed in a column from top to bottom vertically on a wall
  10. wall_horizontal_row - objects are placed in a row from left to right vertically on a wall
  11. wall_grid - objects are placed in a grid orderly and vertically on a wall (when in doubt between grid and wall_grid, choose wall_grid if the objects are placed on a wall)
  12. letter - objects are placed to form a letter of the alphabet
  13. rectangular_perimeter - objects are placed around the perimeter of a rectangular shape facing inward
  If the motif type is letter, concatenate the letter to the end of the motif type (e.g., letter_A, letter_B, etc.)
  Respond with the name of the motif type only.
\end{minted}

\subsection{Prompts to Generate \motifprogcap from \naiveprogcap}
\label{sec:prompts-motif-program}

When generating the \motifprog, our observational prompts instruct the LLM to make high-level observations about the following:
\begin{itemize}
\item the number and type of objects in the arrangement.
\item the relative displacements between objects.
\item spatial patterns such as symmetries.
\item spatial patterns along the three coordinate axes.
\end{itemize}

\begin{minted}[breaklines, breakafter=d, fontsize=\scriptsize]{yaml}
# Observation
optimize_highlevel_count: >-
  Below is a program about a spatial motif of "<DESCRIPTION>".
  Describe how many object types and how many are there for each type.
  Respond with a json-like text structure with the object types as keys and the counts as values.
  Here is the program:
  ```python
  <PROGRAM>
  ```
optimize_highlevel_general_pattern: >-
  Can you observe any pattern in this motif of objects about "<DESCRIPTION>"? (Look for common sense things like repeating layout, symmetry, etc.)
  Respond with a detailed description of the pattern you observed in text.
optimize_highlevel_xyz_pattern: >-
  Can you observe any pattern in the x, y, z coordinates of the objects in this motif?
  Respond with a detailed description of the pattern you observed in text.
optimize_highlevel_xyz_displacements: >-
  Can you find the relative displacements of each object to the next object in the x, y, z dimensions?
  Base on these displacement values, group the objects into different subgroups.
  The goal is to disentangle the overall motif into smaller, self-contained spatial patterns.
  Assign each object to a subgroup such that the spatial pattern in the subgroup can easily be expressed by some programming constructs.
  Each object should belong to exactly one subgroup.
# -------------------------------------------
# Generation
optimize_lowlevel: >-
  Take a deep breath and follow the steps below carefully.
  Using these observations you made, modify the original program such that the code structure represents the spatial pattern.
  Here are the requirements for this task:
  1. Do not create any functions
  2. Store all created objects in a list named `objs`
  3. The new program should represent the same scene as in the original program with the same number of objects.
  4. The objects do not need to be in the same order as the original program, instead, opt for a more structured or natural order.
  5. Do not list all object coordinates, rotation angles, or sizes in a list and acess them in a loop using index.
     This is not acceptable as the spatial pattern is lost.
     Avoid doing this at all costs.
     Use loops, arithmetic operators, if-else statements, and other programming constructs to encapsulate the patterns.
  6. The coordinates and sizes can be slightly different from the original program, as long as the overall motif remains the same.
  Please respond with code only.
# -------------------------------------------
# Validation
validate_naive_listing: >-
  Below is a program that describes a spatial motif of objects.
  Please analyze the program for the following criterion:
  The program should define and manipulate the objects and their properties without using lists of specific object attributes for individual objects.
  It is okay to store common values in variables and reuse them (e.g., dimensions, common properties).
  However, the program should use programming constructs and arithmetic operations (like loops or calculations) to define and manipulate the positional and orientational properties of objects, rather than explicitly listing them.
  For example, if there are four objects, the program should not have a list of four (x, y, z) coordinates or four rotation angles for each object.
  Exclude the `objs` list from the analysis.
  Answer in a json format:
  {'valid': 'yes' or 'no', 'variable_names': a list of the names of the variables that violated the criterion}
  Briefly explain your reasoning in two to three sentences after the json-like text structure.
  Here is the program:
  ```python
  <PROGRAM>
  ```
# -------------------------------------------
# Feedback
optimize_lowlevel_feedback_syntax: >-
  I could not run the program you provided.
  <FEEDBACK>
  Please try again.
optimize_lowlevel_feedback_naive_listing: >-
  Your program creates an motif that accurately recreates the one in the original program.
  However, you listed some attributes of all objects in a list and accesses them in a loop.
  <FEEDBACK>
  Please remove such statements and use loops, arithmetic operators, if-else statements to capture the object attribute patterns
optimize_lowlevel_feedback_num_objs: >-
  The number of objects in the original program does not match the number of objects in your program.
  <FEEDBACK>
  Please try again.
optimize_lowlevel_feedback_centroids: >-
  Objects in the original program are not accurately placed in your program.
  <FEEDBACK>
  Please try again.
optimize_lowlevel_feedback_bounding_boxes: >-
  Objects in the original program are not accurately created in your program.
  <FEEDBACK>
  The objects in your program either have incorrect dimensions or are not oriented correctly.
  Please try again.
\end{minted}

\subsection{Prompts to Generate \metaprogcap from \motifprogcap}
\label{sec:prompts-meta-program}

When generating the \metaprog, our observational prompts instruct the LLM to make high-level observations about the following:
\begin{itemize}
\item commonalities between the motif-programs.
\item differences between the motif-programs.
\end{itemize}

To ensure the generated meta-program is robust and re-useable, we use reasoning prompts to instruct the LLM to consider:
\begin{itemize}
\item reasons why the motif-programs all belong to the same motif
\item what arguments should be extracted
\item what structure should the meta-program function have
\end{itemize}

\begin{minted}[breaklines, breakafter=d, fontsize=\scriptsize]{yaml}
# Observation
generalize_high_level_commonalities: >-
  Take a deep breath and follow the steps below carefully.
  Here are "<NUM_PROGRAMS>" programs that describe the same type of spatial motif: "<MOTIF_TYPE>".
  Each program may have different objects and different coordinates, but they all fall under the same type of spatial motif.
  Here are the programs and the spatial motifs they create:
  <ALL_PROGRAMS>
  For each program, analyze and observe its structure and how the structure coorelates with the spatial motif it creates.
  Pay extra attention to how the objects are placed relative to each other.
  Can you identify the commonalities among these programs?
  If there is only one program, hypothesize what the commonalities could be if there were multiple programs of the same type.
  For the alphabet_letter motif type, only consider the commonalities among the programs that form the same letter.
  Respond with a detailed description of the pattern you observed in text.
generalize_high_level_differences: >-
  What are the differences among these programs?
  Specifically, how are the objects placed differently in each program?
  For each program, observe and describe the differences in the program structure that lead to the different spatial motifs.
  If there is only one program, hypothesize how the program could be written differently to create a different spatial motif of the same type.
  Your hypothesized programs should not deviate from the concrete, non-hypothetical programs you analyzed, and should focus on the number of objects, their positions, rotations, and sizes as the main axes of variation.
  The hypothesized programs should be close to the original program in terms of the spatial motif they create (e.g., a stack of 3 objects instead of 4, or a row of 4 objects instead of 3, etc.)
  Respond with a detailed description of the differences of each program in text.
  Refer to the programs by their order in the previous prompt.
# -------------------------------------------
# Reasoning
generalize_high_level_motif_reason: >-
  Despite these differences, these programs all fall under the same type of spatial motif: "<MOTIF_TYPE>".
  Here is a list of all available motif types:
  1. stack - multiple objects of the same type are placed orderly on top of each other
  2. pile - objects are placed on top of each other but not in an orderly manner (when in doubt between stack and pile, choose pile if the description contains the word "pile" explicitly)
  3. row - objects are placed next to each other in a row
  4. grid - objects are placed orderly in a grid, like a chessboard
  5. left_of - one object placed to the left of another object
  6. in_front_of - one object placed in front of another object
  7. on_top - one object placed on top of another object
  8. surround - objects are placed around a central object in a circular manner
  9. wall_vertical_column - objects are placed in a column from top to bottom vertically on a wall
  10. wall_horizontal_row - objects are placed in a row from left to right vertically on a wall
  11. wall_grid - objects are placed in a grid orderly and vertically on a wall (when in doubt between grid and wall_grid, choose wall_grid if the objects are placed on a wall)
  12. letter - objects are placed to form a letter of the alphabet
  13. rectangular_perimeter - objects are placed around the perimeter of a rectangular shape facing inward
  Based on your observations, what are the reasons that these programs fall under the same type of spatial motif?
  Also, identify the closest incorrect motif type and explain why it is not the correct motif type.
  Respond with a detailed description of the reasons in text.
generalize_low_level_arguments: >-
  With this conversation in mind, you are now tasked to write a python function (a meta-program) that can be used to create various spatial motifs of "<MOTIF_TYPE>".
  A user should be able to call this meta-program with different arguments to create different spatial motifs of this type.
  Using the provided programs as examples, the meta-program should be able to create any spatial motif of this type with slight variations in the number of objects and their attributes.
  Apart from the objects in the examples, the meta-program should be able to create new spatial motifs of different object types and attributes.
  Use your analysis on the commonalities and differences among the example programs to guide you in writing the meta-program.
  Keep in mind that the final meta-program should be able to recreate the given example spatial motifs as closely as possible.
  Avoid taking plain lists of object attributes, such as coordinates, sizes, and rotations, as arguments, as this defeats the purpose of using a meta-program to capture fundamental spatial patterns.
  I will guide you through the process of writing the function step by step.
  First, what are the arguments that the meta-program should take to create a spatial motif of this type?
  Respond with a detailed description of the arguments in text.
generalize_low_level_structure: >-
  You have identified the arguments needed for the meta-program.
  Now, think about the structure of the meta-program.
  How should the meta-program be structured to generalize the spatial motif of "<MOTIF_TYPE>"?
  Aim to encapsulate the commonalities and differences among the example programs in the structure of the meta-program.
  Think about how the arguments should be used in the meta-program to create the spatial motif.
  What programming constructs should be used to encapsulate the commonalities and differences among the example programs?
  Respond with a detailed description of the structure of the meta-program in text.
# -------------------------------------------
# Generation
generalize_low_level: >-
  Take a deep breath and follow the steps below carefully.
  You have identified the arguments and the structure of the meta-program.
  Now, write the meta-program that can create various spatial motifs of "<MOTIF_TYPE>".
  Here is a meta-program of the same motif type you created in the past:
  ```python
  <PAST_META_PROGRAM>
  ```
  You can use this as a reference to write the new meta-program, but you can also write it from scratch.
  Refer to the arguments you identified and the structure you described to write the meta-program.
  Remember that the meta-program should be able to recreate the spatial motifs of the example programs as closely as possible.
  Avoid taking plain lists of object attributes, such as coordinates, sizes, and rotations, as arguments, as this defeats the purpose of using a meta-program to capture fundamental spatial patterns.
  Include a docstring at the beginning of the meta-program to describe the purpose of the function and the arguments it takes.
  Use comments to explain your code step by step.
  Please respond with code only.
# -------------------------------------------
# Validation
generalize_low_level_batch_recreate: >-
  Using the meta-program you wrote, what are the function calls needed to recreate each of the example programs you analyzed previously?
  Write one function call for each example program, that recreates the spatial motif of the program as closely as possible.
  Ignore the programs you hypothesized and only write function calls for the example programs I provided.
  Response with a json-like text structure with the example program order integers as keys and the function calls as values.
  For example, {"1": "function_call(program1_args)", "2": "function_call(program2_args)", ...}
# -------------------------------------------
# Feedback
generalize_low_level_feedback: >-
  The meta-program you wrote could not recreate the spatial motif of the example programs as closely as possible.
  Here are the feedback(s) for the example program(s) that were not recreated accurately:
  <FEEDBACK>
  The issue(s) may be related to the meta-program structure or the arguments used in the function calls.
  Please analyze the feedback and modify the meta-program if necessary.
  Remember that the meta-program should be able to recreate the spatial motifs of the example programs as closely as possible.
  Also, remember to edit the docstring and comments in the meta-program to reflect the changes you made.
  Then, respond with the code of the modified meta-program only.
  If you think the meta-program is correct, and the function calls are incorrect, then you can respond with the same meta-program code.
  I will ask you to provide the function calls again after this where you can correct the function calls.
generalize_refine_comments: >-
  You have successfully written the meta-program to generalize the spatial motif of "<MOTIF_TYPE>".
  The latest function calls you provided are able to recreate the spatial motifs of the example programs.
  Please update the docstring and comments in the meta-program to provide as much information as possible about the expected arguments.
  If the program contains well-defined parts that can be explained, include comments to describe these parts - how they contribute to the spatial motif and how they differ from other parts.
  At inference time, the meta-program will be the sole source of information for recreating this spatial motif.
  Therefore, these comments will be crucial for understanding how the meta-program should be used.
  Include at least one example function call in the comments to show how the meta-program should be called.
  Please respond with code only.
\end{minted}

\subsection{Inference Prompts}
\label{sec:prompts-inference}

\begin{minted}[breaklines, breakafter=d, fontsize=\scriptsize]{yaml}
inference: >-
  Here is a meta-program that generalizes a spatial arrangement of type "<MOTIF_TYPE>":
  ```python
  <META_PROGRAM>
  ```
  And here is a description of a spatial motif of the same type:
  <DESCRIPTION>
  Your task is to call the meta-program with the necessary arguments to recreate the spatial motif described in the description as closely as possible.
  Read the docstring and comments in the meta-program to understand how to use it.
  Refer to the example function call in the meta-program documentation to understand how the meta-program should be called, if available.
  Use common sense to infer the arguments for ambiguous arguments, such as object dimensions, positions, and rotations.
  When in doubt, refer back to the example function call in the meta-program documentation.
  Ensure the arguments make the new spatial motif physically possible without intersecting or floating objects and make the objects contact tightly, if applicable.
  I will run a postprocessing step to refine the spatial motif after you provide the function call to me.
  Remember that x, y, z dimensions are in meters and rotation angles are in degrees.
  +x is to the right, +y is up, and +z is front (towards the viewer).
  Most object rotations hence are around the y-axis, as it is the vertical axis.
  Assume the objects are large so make sure they are spaced out enough to not intersect.
  When placing objects, consider how human would organize them in real life.
  For example, large objects are usually placed at the bottom and smaller objects on top.
  Please respond with code - the function call with the inferred arguments only.
inference_feedback: >-
  I could not run the meta-program using the function call you provided.
  <FEEDBACK>
  Please try again.
# -------------------------------------------
retrieval_mesh_rotations: >-
  I am retrieving meshes of objects from a human-authored object dataset to instantiate a spatial motif of "<DESCRIPTION>".
  The labels of the objects are "<OBJECT_LABELS>".
  How likely would be the retrieved mesh in the correct orientation?
  Use common sense knowledge to reason about this.
  Consider whether the objects are commonly oriented in both upright and sideways positions in real life.
  If there are multiple common orientations, then the retrieved mesh is likely to be in the incorrect orientation.
  Respond in a json-like text structure with the object labels as keys and two probabilities ("correct", "incorrect") as values l
  ```json
  {"obj1": {"correct": X, "incorrect": 1-X}, "obj2": ...}
  ```
  where X is between 0 and 1.
  The two probabilities "correct", "incorrect" indicate how likely the retrieved mesh would be in the correct orientation for the motif.
  For objects that are hanged on a wall, always consider the mesh would be in the incorrect orientation.
  Include a brief two to three sentences explanation for each object after the json-like text structure.
spatial_optimization_touch: >-
  I have instantiated a spatial motif of "<DESCRIPTION>" using meshes retrieved from a human-authored object dataset.
  How common is it for the individual objects to be placed in tight contact with each other in this motif?
  Consider whether this motif describes a decorative arrangement or a functional arrangement.
  A decorative arrangement is one that is intentionally designed to be arranged in a specific way for aesthetic purposes.
  If the motif is a decorative arrangement, then touching is less common.
  Otherwise, use common sense to reason about whether the objects are typically placed in tight contact with each other in real life.
  Answer using percentage to indicate how common each version is.
  Answer in a json-like text structure with two probabilities like below:
  ```json
  {"touch": X, "no_touch": 1-X}
  ```
  where X, between 0 and 1, is the probability of the version fitting the common sense definition.
  Briefly explain your reasoning in two to three sentences after the json-like text structure.
# -------------------------------------------
wnsynsetkeys: >-
  Here is a list of wordnet synset keys.
  <WNSYNSETKEYS>
  What is the best key for the following object labels?
  <OBJECT_LABELS>
  Response with a json-like list of synset keys in the same order as the object labels like below:
  ```json
  {"wnsynsetkeys": ["key1", "key2", "key3"]}
  ```
  Use "none" if none of the keys are suitable for an object.
# -------------------------------------------
invalid_response: >-
  <FEEDBACK>
  Please try again.
\end{minted}

\subsection{Ablation Prompts}
\label{sec:prompts-ablation}

Simple prompts for generating a meta-program directly from a description of a spatial motif (without example arrangement), directly from a set of motif programs (without any reasoning), and through in-context learning.
\begin{minted}[breaklines, breakafter=d, fontsize=\scriptsize]{yaml}
ablation_direct: >-
  Using the DSL, write a program about a spatial motif of "<DESCRIPTION>".
  Please respond with code only.
  Do not define any function.
  Your code should be a list of statements that can be executed sequentially and directly without any modifications or extra function calls.
  Please respond with code only.
ablation_generalize: >-
  Here are different programs that describe the same type of spatial motif: "<MOTIF_TYPE>".
  Each program may have different objects and different coordinates, but they all fall under the same type of spatial motif.
  Here are the programs and the spatial motifs they create:
  <ALL_PROGRAMS>
  Now, write a meta-program that can generalize the spatial motif of this type using the provided programs as examples.
  The meta-program should be able to create any spatial motif of this type with slight variations in the number of objects and their attributes.
  Identify the arguments that the meta-program should take and the structure of the meta-program to generalize the spatial motif.
  Put the arguemnts in the function signature and write the meta-program as a callable function.
  The function should return the list of objects created in the spatial motif.
  Here is a meta-program of the same motif type you created in the past:
  ```python
  <PAST_META_PROGRAM>
  ```
  You can use this as a reference to write the new meta-program, but you can also write it from scratch.
  Include a docstring at the beginning of the meta-program to describe the purpose of the function and the arguments it takes.
  Use comments to explain your code step by step.
  Please respond with code only.
ablation_incontext: >-
  Using the DSL, write a program about a spatial motif of "<DESCRIPTION>".
  Here are programs that describe the same type of spatial motif: "<MOTIF_TYPE>".
  Use these programs as examples to guide you in writing the program:
  <ALL_PROGRAMS>
  Please respond with code only.
\end{minted}

%% file: supp/session.tex
\section{Example Session}
\label{sec:supp-session}

We show an example session generating meta-program given arrangement corresponding to ``a stack of seven plates''.
We provide both the prompts to the LLM and the responses from the LLM for different phases of learning to generate the meta-program.
Note that the session is fully automatic and does not require any user intervention.

\subsection{\naiveprogcap Extraction}

\noindent Initial \naiveprog extracted directly from the example arrangement (does not require a LLM).

\begin{minted}[breaklines, breakafter=d, fontsize=\scriptsize]{python}
# Description: a stack of seven plates
# Naive program extracted from input arrangment
objs = []
obj_1_half_size = [0.08909, 0.0143, 0.08853]
obj_1_centroid = [0.0, 0.0, 0.0]
obj_1 = create('plate', obj_1_half_size)
move(obj_1, obj_1_centroid[0], obj_1_centroid[1], obj_1_centroid[2])
objs.append(obj_1)
obj_2_half_size = [0.08909, 0.0143, 0.08853]
obj_2_centroid = [0.0, -0.00757, 0.0]
obj_2 = create('plate', obj_2_half_size)
move(obj_2, obj_2_centroid[0], obj_2_centroid[1], obj_2_centroid[2])
objs.append(obj_2)
obj_3_half_size = [0.08909, 0.0143, 0.08853]
obj_3_centroid = [0.0, -0.01514, 0.0]
obj_3 = create('plate', obj_3_half_size)
move(obj_3, obj_3_centroid[0], obj_3_centroid[1], obj_3_centroid[2])
objs.append(obj_3)
obj_4_half_size = [0.08909, 0.0143, 0.08853]
obj_4_centroid = [0.0, -0.02272, -0.0]
obj_4 = create('plate', obj_4_half_size)
move(obj_4, obj_4_centroid[0], obj_4_centroid[1], obj_4_centroid[2])
objs.append(obj_4)
obj_5_half_size = [0.08909, 0.0143, 0.08853]
obj_5_centroid = [0.0, -0.03029, -0.0]
obj_5 = create('plate', obj_5_half_size)
move(obj_5, obj_5_centroid[0], obj_5_centroid[1], obj_5_centroid[2])
objs.append(obj_5)
obj_6_half_size = [0.08909, 0.0143, 0.08853]
obj_6 = create('plate', obj_6_half_size)
move(obj_6, obj_6_centroid[0], obj_6_centroid[1], obj_6_centroid[2])
objs.append(obj_6)
obj_7_half_size = [0.08909, 0.0143, 0.08853]
obj_7_centroid = [0.0, -0.04544, 0.0]
obj_7 = create('plate', obj_7_half_size)
move(obj_7, obj_7_centroid[0], obj_7_centroid[1], obj_7_centroid[2])
objs.append(obj_7)
\end{minted}

\subsection{\motifprogcap Generation}

\mypara{High level observations of \naiveprog.}

\noindent Asking LLM to make high level observations of the \naiveprog using prompt \prompt{optimize\_highlevel\_count}
\begin{minted}[breaklines, breakafter=d, fontsize=\scriptsize]{text}
Below is a program about a spatial motif of "<DESCRIPTION>".
Describe how many object types and how many are there for each type.
Respond with a json-like text structure with the object types as keys and the counts as values.
Here is the program:
```python
<PROGRAM>
```
\end{minted}

\noindent Response from LLM (json)
\begin{minted}[breaklines, breakafter=d, fontsize=\scriptsize]{json}
{
  "plate": 7
}
\end{minted}

\noindent Asking LLM to make high level observations of the \naiveprog using prompt \\ \prompt{optimize\_highlevel\_general\_pattern}
\begin{minted}[breaklines, breakafter=d, fontsize=\scriptsize]{text}
Can you observe any pattern in this motif of objects about "<DESCRIPTION>"? (Look for common sense things like repeating layout, symmetry, etc.)
Respond with a detailed description of the pattern you observed in text.
\end{minted}

\noindent Response from LLM (text)
\begin{minted}[breaklines, breakafter=d, fontsize=\scriptsize]{text}
The pattern observed in the "stack of seven plates" can be described as follows:

1. **Object Consistency**: All seven objects in the stack are of the same type, i.e., "plate." They all share identical dimensions, with a half-size of [0.08909, 0.0143, 0.08853].

2. **Vertical Stacking**: The plates are vertically stacked along the y-axis. This observation can be made from the changes in the `y` component of the centroids of each plate.

3. **Even Spacing**: The centroids of the plates are evenly spaced along the negative y-axis. Each subsequent plate is positioned 0.00757 meters below the previous one. This incremental change in the y-coordinate confirms a precise stacking arrangement.

4. **Fixed x and z coordinates**: The `x` and `z` coordinates of the centroids of all the plates remain constant at 0.0. This indicates the stack is perfectly aligned along these axes, ensuring that the stack is upright and not tilted or offset in any horizontal direction.

Given these observations, the motif can be described as a neat, symmetrical vertical stack of seven uniform plates, each spaced evenly from one another.
\end{minted}

\noindent Asking LLM to make high level observations of the \naiveprog using prompt \prompt{optimize\_highlevel\_xyz\_pattern}
\begin{minted}[breaklines, breakafter=d, fontsize=\scriptsize]{text}
Can you observe any pattern in the x, y, z coordinates of the objects in this motif?
Respond with a detailed description of the pattern you observed in text.
\end{minted}

\noindent Response from LLM (text)
\begin{minted}[breaklines, breakafter=d, fontsize=\scriptsize]{text}
The x, y, and z coordinates of the objects in this motif exhibit the following patterns:

### x-coordinate:
- The x-coordinate for the centroids of all plates is 0.0.
- This consistent value indicates that there is no horizontal shifting in the x-direction, meaning all plates are aligned vertically without any horizontal displacement.

### y-coordinate:
- The y-coordinates show a clear pattern with a consistent decrease of 0.00757 meters for each subsequent plate.
- The initial y-coordinate is 0.0 for the first plate, and the y-coordinates for the following plates are:
  - 0.0 for the 1st plate
  - -0.00757 for the 2nd plate
  - -0.01514 for the 3rd plate
  - -0.02272 for the 4th plate
  - -0.03029 for the 5th plate
  - -0.03786 for the 6th plate
  - -0.04544 for the 7th plate
- This indicates that the plates are evenly spaced along the vertical axis (negative y-direction) by a distance of 0.00757 meters.

### z-coordinate:
- The z-coordinate for the centroids of all plates is consistently 0.0.
- This indicates there is no depth displacement along the z-axis, meaning the stack is aligned in the depth direction as well, with no forward or backward shifting.

### Pattern Summary:
- **Vertical Alignment**: All plates are perfectly aligned in the vertical axis (y-axis) with an even spacing.
- **No Horizontal Displacement**: There is no change in the x-coordinate, ensuring horizontal alignment.
- **No Depth Displacement**: There is no change in the z-coordinate, ensuring depth alignment.

In conclusion, the coordinates exhibit a clear pattern of a vertically stacked arrangement with precise spacing and no horizontal or depth displacement, ensuring a neat and uniform stack of plates.
\end{minted}

\noindent Asking LLM to make high level observations of the \naiveprog using prompt \\ \prompt{optimize\_highlevel\_xyz\_displacements}
\begin{minted}[breaklines, breakafter=d, fontsize=\scriptsize]{text}
Can you find the relative displacements of each object to the next object in the x, y, z dimensions?
Base on these displacement values, group the objects into different subgroups.
The goal is to disentangle the overall motif into smaller, self-contained spatial patterns.
Assign each object to a subgroup such that the spatial pattern in the subgroup can easily be expressed by some programming constructs.
Each object should belong to exactly one subgroup.
\end{minted}

\noindent Response from LLM (text)
\begin{minted}[breaklines, breakafter=d, fontsize=\scriptsize]{text}
### Relative Displacements:
The relative displacements of each object to the next object in the stack are as follows:

- **Displacements**:
  - Object 1 to Object 2: (0.0, -0.00757, 0.0)
  - Object 2 to Object 3: (0.0, -0.00757, 0.0)
  - Object 3 to Object 4: (0.0, -0.00757, 0.0)
  - Object 4 to Object 5: (0.0, -0.00757, 0.0)
  - Object 5 to Object 6: (0.0, -0.00757, 0.0)
  - Object 6 to Object 7: (0.0, -0.00757, 0.0)

Each of these displacements is identical, indicating a uniform vertical displacement between consecutive objects.

### Grouping into Subgroups:
Given the uniform relative displacement pattern, the entire set of objects can be grouped into one single subgroup. This is due to the consistent and repetitive nature of the displacement, which can be expressed succinctly using a loop construct in a programming context.

### Subgroup Assignment:
- **Subgroup 1**: All objects (Object 1 to Object 7)

### Expressed Subgroup Pattern:
- The spatial pattern in this subgroup is a uniform vertical displacement of 0.00757 meters along the negative y-axis between consecutive objects.

### Programming Context:
- This can be expressed using a loop to create and move each object in the subgroup with the uniform displacement.
\end{minted}

\mypara{Determining motif type for storing into program library.}

\noindent Asking LLM to classify motif type for the description using prompt \prompt{classify}
\begin{minted}[breaklines, breakafter=d, fontsize=\scriptsize]{text}
Based on your observations, classify the spatial motif of objects "<DESCRIPTION>" into one of the following motif types:
1. stack - multiple objects of the same type are placed orderly on top of each other
2. pile - objects are placed on top of each other but not in an orderly manner (when in doubt between stack and pile, choose pile if the description contains the word "pile" explicitly)
3. row - objects are placed next to each other in a row
4. grid - objects are placed orderly in a grid, like a chessboard
5. left_of - one object placed to the left of another object
6. in_front_of - one object placed in front of another object
7. on_top - one object placed on top of another object
8. surround - objects are placed around a central object in a circular manner
9. wall_vertical_column - objects are placed in a column from top to bottom vertically on a wall
10. wall_horizontal_row - objects are placed in a row from left to right vertically on a wall
11. wall_grid - objects are placed in a grid orderly and vertically on a wall (when in doubt between grid and wall_grid, choose wall_grid if the objects are placed on a wall)
12. letter - objects are placed to form a letter of the alphabet
13. rectangular_perimeter - objects are placed around the perimeter of a rectangular shape facing inward
If the motif type is letter, concatenate the letter to the end of the motif type (e.g., letter_A, letter_B, etc.)
Respond with the name of the motif type only.
\end{minted}

\noindent Response from LLM (text)
\begin{minted}[breaklines, breakafter=d, fontsize=\scriptsize]{text}
stack
\end{minted}

\mypara{Motif program generation.}

\noindent Asking LLM to transform the \naiveprog into a \motifprog based on the observations made by the LLM using the prompt \prompt{optimize\_lowlevel}
\begin{minted}[breaklines, breakafter=d, fontsize=\scriptsize]{text}
Take a deep breath and follow the steps below carefully.
Using these observations you made, modify the original program such that the code structure represents the spatial pattern.
Here are the requirements for this task:
1. Do not create any functions
2. Store all created objects in a list named `objs`
3. The new program should represent the same scene as in the original program with the same number of objects.
4. The objects do not need to be in the same order as the original program, instead, opt for a more structured or natural order.
5. Do not list all object coordinates, rotation angles, or sizes in a list and acess them in a loop using index.
 This is not acceptable as the spatial pattern is lost.
 Avoid doing this at all costs.
 Use loops, arithmetic operators, if-else statements, and other programming constructs to encapsulate the patterns.
6. The coordinates and sizes can be slightly different from the original program, as long as the overall motif remains the same.
Please respond with code only.
\end{minted}

\noindent Response from LLM (python)
\begin{minted}[breaklines, breakafter=d, fontsize=\scriptsize]{python}
objs = []
half_size = [0.08909, 0.0143, 0.08853]
initial_y = 0.0
displacement_y = -0.00757

for i in range(7):
    y_position = initial_y + i * displacement_y
    obj = create('plate', half_size)
    move(obj, 0.0, y_position, 0.0)
    objs.append(obj)
\end{minted}

\mypara{Motif program validation.}

\noindent Asking LLM to validate the \motifprog using the prompt \prompt{validate\_naive\_listing}.
\begin{minted}[breaklines, breakafter=d, fontsize=\scriptsize]{text}
Below is a program that describes a spatial motif of objects.
Please analyze the program for the following criterion:
The program should define and manipulate the objects and their properties without using lists of specific object attributes for individual objects.
It is okay to store common values in variables and reuse them (e.g., dimensions, common properties).
However, the program should use programming constructs and arithmetic operations (like loops or calculations) to define and manipulate the positional and orientational properties of objects, rather than explicitly listing them.
For example, if there are four objects, the program should not have a list of four (x, y, z) coordinates or four rotation angles for each object.
Exclude the `objs` list from the analysis.
Answer in a json format:
{'valid': 'yes' or 'no', 'variable_names': a list of the names of the variables that violated the criterion}
Briefly explain your reasoning in two to three sentences after the json-like text structure.
Here is the program:
```python
<PROGRAM>
```
\end{minted}

\noindent Response from LLM (json)
\begin{minted}[breaklines, breakafter=d, fontsize=\scriptsize]{json}
{
  "valid": "yes",
  "variable_names": []
}
\end{minted}

\noindent All validation checks were passed after iteration 1.
The program provided by the LLM is the stored in the program library with the classified motif type of \prompt{stack}.

\noindent If the validation did not pass, then the \prompt{optimize\_lowlevel\_feedback} prompts are used to refine the program based on the validation test that didn't pass.

\subsection{\metaprogcap Generation}

\noindent Asking LLM to make observations of \emph{commonalities} in a set of \motifprog with motif type \prompt{stack} using the prompt \\ \prompt{generalize\_high\_level\_commonalities}

\begin{minted}[breaklines, breakafter=d, fontsize=\scriptsize]{text}
Take a deep breath and follow the steps below carefully.
Here are "<NUM_PROGRAMS>" programs that describe the same type of spatial motif: "<MOTIF_TYPE>".
Each program may have different objects and different coordinates, but they all fall under the same type of spatial motif.
Here are the programs and the spatial motifs they create:
<ALL_PROGRAMS>
For each program, analyze and observe its structure and how the structure coorelates with the spatial motif it creates.
Pay extra attention to how the objects are placed relative to each other.
Can you identify the commonalities among these programs?
If there is only one program, hypothesize what the commonalities could be if there were multiple programs of the same type.
For the alphabet_letter motif type, only consider the commonalities among the programs that form the same letter.
Respond with a detailed description of the pattern you observed in text.
\end{minted}

\noindent Response from LLM (text)
\begin{minted}[breaklines, breakafter=d, fontsize=\scriptsize]{text}
Based on the analysis of Program 1, which describes a "stack" spatial motif, we can identify several commonalities that characterize this type of spatial arrangement. Here are the observed commonalities and patterns:

### Commonalities of the Stack Motif:

1. **Object Type Consistency**:
   - All objects in the motif are of the same type. In this case, all objects are identified as 'plate'. This consistency ensures uniformity in the appearance and arrangement of the stack.

2. **Uniform Object Dimensions**:
   - The objects in the stack all have the same dimensions, represented by the `half_size` parameter. This uniformity in size contributes to the regular and orderly appearance of the stack.

3. **Vertical Alignment**:
   - The objects are aligned vertically along the y-axis. This is characteristic of a stack, where objects are placed directly on top of one another in a straight line.

4. **Even Spacing**:
   - There is a consistent vertical displacement between consecutive objects in the stack. In Program 1, the displacement is uniform with a negative y-coordinate increment of -0.00757 meters. This regular spacing ensures that the stack maintains a tidy and organized appearance.     

5. **Fixed Horizontal and Depth Positions**:
   - The x and z coordinates of the objects remain constant. In Program 1, these coordinates are both 0.0 for all objects, indicating that there is no horizontal or depth displacement. This fixed positioning along the x and z axes maintains the straight vertical alignment characteristic of a stack.

### Hypothesized Commonalities for Multiple Programs:
If there were multiple programs describing the "stack" motif, we might observe similar commonalities:

1. **Object Type**:
   - Multiple programs could feature different types of objects (e.g., books, boxes), but within each program, the objects would be consistently of the same type.

2. **Uniform Dimensions**:
   - Even if different programs use different objects, the dimensions of the objects within each program would remain uniform.

3. **Vertical Spacing**:
   - The vertical displacement between objects might vary between programs, but within each program, the displacement would be consistent and regular.

4. **Vertical Alignment**:
   - Each program would align the objects vertically along the y-axis, maintaining this key characteristic of a stack.

5. **Fixed x and z Coordinates**:
   - While different programs might place the stack at different positions in the world, within each program, the x and z coordinates would be fixed for all stacked objects to enforce the vertical alignment.

### Conclusion:
The commonalities observed in the "stack" motif emphasize uniformity in object type and size, regular vertical spacing, and maintaining a consistent vertical alignment without horizontal or depth displacement. These patterns ensure that the spatial arrangement of objects reflects a neat and orderly stack.
\end{minted}

\noindent Asking LLM to make observations of \emph{differences} in a set of \motifprog with motif type \prompt{stack} using the prompt \\ \prompt{generalize\_high\_level\_differences}
\begin{minted}[breaklines, breakafter=d, fontsize=\scriptsize]{text}
What are the differences among these programs?
Specifically, how are the objects placed differently in each program?
For each program, observe and describe the differences in the program structure that lead to the different spatial motifs.
If there is only one program, hypothesize how the program could be written differently to create a different spatial motif of the same type.
Your hypothesized programs should not deviate from the concrete, non-hypothetical programs you analyzed, and should focus on the number of objects, their positions, rotations, and sizes as the main axes of variation.
The hypothesized programs should be close to the original program in terms of the spatial motif they create (e.g., a stack of 3 objects instead of 4, or a row of 4 objects instead of 3, etc.)
Respond with a detailed description of the differences of each program in text.
Refer to the programs by their order in the previous prompt.
\end{minted}

\noindent Response from LLM (text)
\begin{minted}[breaklines, breakafter=d, fontsize=\scriptsize]{text}
Since we have only one concrete program to analyze, let's hypothesize how it could be written differently to create different variations of the same spatial motif, "stack." We will focus on varying the number of objects, their positions, and possibly other factors such as rotations (though rotations might be less relevant for the stack motif).

### Hypothesized Variations:

1. **Changing the Number of Objects**:
   - Instead of a stack of seven plates, we can create a stack with a different number of plates. For instance:

   ```python
   objs = []
   half_size = [0.08909, 0.0143, 0.08853]
   initial_y = 0.0
   displacement_y = -0.00757

   for i in range(5):  # Stack of 5 plates instead of 7
       y_position = initial_y + i * displacement_y
       obj = create('plate', half_size)
       move(obj, 0.0, y_position, 0.0)
       objs.append(obj)
   ```
   - **Difference**: The main difference is the reduction in the number of objects from 7 to 5, leading to a shorter stack.

2. **Changing the Vertical Displacement**:
   - By modifying the vertical displacement, we can create a stack with different spacing between objects:

   ```python
   objs = []
   half_size = [0.08909, 0.0143, 0.08853]
   initial_y = 0.0
   displacement_y = -0.01  # Increased spacing between plates

   for i in range(7):
       y_position = initial_y + i * displacement_y
       obj = create('plate', half_size)
       move(obj, 0.0, y_position, 0.0)
       objs.append(obj)
   ```
   - **Difference**: The increased vertical displacement creates a stack with more space between each plate compared to the original program.

3. **Introducing Randomness**:
   - Adding a slight randomness to the displacement could simulate a less orderly but still essentially vertical stack:

   ```python
   import random

   objs = []
   half_size = [0.08909, 0.0143, 0.08853]
   initial_y = 0.0
   displacement_y = -0.00757

   for i in range(7):
       random_offset = random.uniform(-0.001, 0.001)
       y_position = initial_y + i * displacement_y + random_offset
       obj = create('plate', half_size)
       move(obj, 0.0, y_position, 0.0)
       objs.append(obj)
   ```
   - **Difference**: Adding a slight random offset to the y-position for each plate introduces minor variations, making it look less perfect while still maintaining the overall stack structure.

4. **Using Different Object Sizes**:
   - Introducing variety in object sizes while maintaining the stack structure:

   ```python
   objs = []
   initial_y = 0.0
   displacement_y = -0.00757

   for i in range(7):
       half_size = [0.08909, 0.0143 + i * 0.002, 0.08853]  # Incremental increase in size
       y_position = initial_y + i * displacement_y
       obj = create('plate', half_size)
       move(obj, 0.0, y_position, 0.0)
       objs.append(obj)
   ```
   - **Difference**: Varying the size slightly with each subsequent object while maintaining the stack adds a gradient or stepped effect to the stack without deviating from the vertical alignment.

### Summary of Differences:
- **Number of Objects**: Varying the quantity alters the height and the visual appearance of the stack.
- **Vertical Displacement**: Adjusting the spacing between objects changes the density/compression of the stack.
- **Randomness**: Introducing slight variations in positioning to make the stack appear less mechanically perfect.
- **Object Sizes**: Varying object sizes can add a dynamic look while retaining the core stack motif.

Each of these variations adheres to the overarching "stack" motif but introduces subtle differences in how the objects are placed relative to each other. These variations maintain the essence of a vertical arrangement while exploring different aspects of spatial positioning and object properties.
\end{minted}

\mypara{Meta-program reasoning.}

\noindent Asking LLM to explain why all \motifprog belong to the motif type \prompt{stack} using the prompt \prompt{generalize\_high\_level\_motif\_reason}

\begin{minted}[breaklines, breakafter=d, fontsize=\scriptsize]{text}
Despite these differences, these programs all fall under the same type of spatial motif: "<MOTIF_TYPE>".
Here is a list of all available motif types:
1. stack - multiple objects of the same type are placed orderly on top of each other
2. pile - objects are placed on top of each other but not in an orderly manner (when in doubt between stack and pile, choose pile if the description contains the word "pile" explicitly)
3. row - objects are placed next to each other in a row
4. grid - objects are placed orderly in a grid, like a chessboard
5. left_of - one object placed to the left of another object
6. in_front_of - one object placed in front of another object
7. on_top - one object placed on top of another object
8. surround - objects are placed around a central object in a circular manner
9. wall_vertical_column - objects are placed in a column from top to bottom vertically on a wall
10. wall_horizontal_row - objects are placed in a row from left to right vertically on a wall
11. wall_grid - objects are placed in a grid orderly and vertically on a wall (when in doubt between grid and wall_grid, choose wall_grid if the objects are placed on a wall)
12. letter - objects are placed to form a letter of the alphabet
13. rectangular_perimeter - objects are placed around the perimeter of a rectangular shape facing inward
Based on your observations, what are the reasons that these programs fall under the same type of spatial motif?
Also, identify the closest incorrect motif type and explain why it is not the correct motif type.
Respond with a detailed description of the reasons in text.
\end{minted}

\noindent Response from LLM (text)
\begin{minted}[breaklines, breakafter=d, fontsize=\scriptsize]{text}
### Reasons These Programs Fall Under the "Stack" Motif:

1. **Object Consistency**:
   - All objects in each program are of the same type. For example, in the provided program, all objects are 'plate'. This consistency ensures that the overall structure is uniform.

2. **Vertical Alignment**:
   - Objects are placed in a vertical line with consistent alignment along the y-axis. Each object is positioned directly above or below the previous one, creating an orderly vertical configuration.

3. **Uniform Spacing**:
   - There is a consistent vertical displacement between consecutive objects. This regular spacing further reinforces the orderly arrangement typical of a stack.

4. **Fixed x and z Coordinates**:
   - The x and z coordinates of all objects remain constant, ensuring that there is no horizontal or depth displacement. This fixed positioning along the x and z axes maintains the strict vertical alignment required for a stack.

5. **Orderliness**:
   - The objects are placed in an orderly manner, directly on top of each other. There is no randomness or displacement that would suggest disorganized placement.

### Closest Incorrect Motif Type:

The closest incorrect motif type is **"pile"**.

### Why "Pile" is Not Correct:

1. **Disorganized Placement**:
   - The "pile" motif is characterized by objects placed on top of each other but not in an orderly manner. This typically implies some degree of randomness or irregular displacement among the objects.
   - In the analyzed and hypothesized programs, the placement of objects is precise, with no randomness introduced that would suggest a disorganized pattern. Even when slight randomness is added in one of the hypothesized variations, the overall structure remains far too regular and ordered to be considered a pile.

2. **Lack of Explicit Disorder**:
   - Descriptions of the motif type "pile" often explicitly include the word "pile," indicating a conscious choice to describe a disorganized arrangement.
   - The provided programs have a clear, structured approach to object placement, consistently using precise coordinates and displacements to maintain order, which does not align with the definition of a "pile."

### Summary:

These programs fall under the "stack" motif because they involve multiple objects of the same type placed in an orderly vertical arrangement with consistent spacing and alignment. The closest incorrect motif type, "pile," is not appropriate because it implies disorganized placement, which contradicts the precise and orderly nature of the stack motif observed in the programs. The consistency in object type, uniform spacing, vertical alignment, and lack of random disorder are key reasons for classifying these programs as "stack" rather than any other motif type.
\end{minted}

\noindent Asking LLM to reason about how to generate a \metaprog for the motif type \prompt{stack} by first identifying the \emph{arguments} using the prompt \prompt{generalize\_low\_level\_arguments}

\begin{minted}[breaklines, breakafter=d, fontsize=\scriptsize]{text}
With this conversation in mind, you are now tasked to write a python function (a meta-program) that can be used to create various spatial motifs of "<MOTIF_TYPE>".
A user should be able to call this meta-program with different arguments to create different spatial motifs of this type.
Using the provided programs as examples, the meta-program should be able to create any spatial motif of this type with slight variations in the number of objects and their attributes.
Apart from the objects in the examples, the meta-program should be able to create new spatial motifs of different object types and attributes.
Use your analysis on the commonalities and differences among the example programs to guide you in writing the meta-program.
Keep in mind that the final meta-program should be able to recreate the given example spatial motifs as closely as possible.
Avoid taking plain lists of object attributes, such as coordinates, sizes, and rotations, as arguments, as this defeats the purpose of using a meta-program to capture fundamental spatial patterns.
I will guide you through the process of writing the function step by step.
First, what are the arguments that the meta-program should take to create a spatial motif of this type?
Respond with a detailed description of the arguments in text.
\end{minted}

\noindent Response from LLM (text)
\begin{minted}[breaklines, breakafter=d, fontsize=\scriptsize]{text}
To create a versatile meta-program that can generate various spatial motifs of the "stack" type, we need to carefully consider the arguments that the function should accept. These arguments should allow for flexibility in creating different stacks while adhering to the core patterns observed in the provided examples. Here are the key arguments:

### Arguments for the Meta-Program:

1. **Object Type (`label`)**:
   - The type of object to be created, which will determine the label for each object in the stack (e.g., 'plate', 'box', etc.).

2. **Number of Objects (`num_objects`)**:
   - The total number of objects to be included in the stack. This allows flexibility in defining the height or length of the stack by varying the number of constituent objects.

3. **Initial Position (`initial_position`)**:
   - A list or tuple representing the initial world coordinates (x, y, z) for the bottom or starting object in the stack.

4. **Vertical Displacement (`displacement_y`)**:
   - The consistent vertical distance between consecutive objects in the stack. This will determine the regular spacing between the objects along the y-axis.

5. **Object Size (`half_size`)**:
   - The dimensions of each object in the stack, represented as a list of three floats [half_width, half_height, half_depth]. This ensures that all objects in the stack have uniform sizes.

6. **Random Offset (`random_offset`)** (optional):
   - An optional argument specifying the range of random variation in the vertical displacement for each object. This can add slight randomness to the positioning, creating a less mechanically perfect stack if desired.

### Additional Considerations:
- **Rotation**: While rotations might be less relevant for a perfectly vertical stack, we could include an optional argument for rotating objects around one or more axes, though this will be kept simple and optional.
- **Axis of Stack**: For future extensibility, we might consider allowing the stack to be along different axes, though for now, we focus on the y-axis for simplicity.

### Summary of Arguments:
Here is a refined list:
1. `label (str)`: Type of object to create (e.g., 'plate').
2. `num_objects (int)`: Number of objects in the stack.
3. `initial_position (list[float])`: Initial position [x, y, z] of the first object.
4. `displacement_y (float)`: Vertical displacement between objects.
5. `half_size (list[float])`: Dimensions of each object as [half_width, half_height, half_depth].
6. `random_offset (float, optional)`: Range for random variation in vertical displacement.

The next step will be to proceed to writing the meta-program using these arguments to generate the desired spatial motifs.
\end{minted}

\noindent Asking LLM to reason about how to generate a \metaprog for the motif type \prompt{stack} by first identifying the \emph{function structure} using the prompt \prompt{generalize\_low\_level\_structure}
\begin{minted}[breaklines, breakafter=d, fontsize=\scriptsize]{text}
You have identified the arguments needed for the meta-program.
Now, think about the structure of the meta-program.
How should the meta-program be structured to generalize the spatial motif of "<MOTIF_TYPE>"?
Aim to encapsulate the commonalities and differences among the example programs in the structure of the meta-program.
Think about how the arguments should be used in the meta-program to create the spatial motif.
What programming constructs should be used to encapsulate the commonalities and differences among the example programs?
Respond with a detailed description of the structure of the meta-program in text.
\end{minted}

\noindent Response from LLM (text)
\begin{minted}[breaklines, breakafter=d, fontsize=\scriptsize]{text}
### Structure of the Meta-Program:

To generalize the spatial motif of "stack," the meta-program needs to encapsulate the commonalities and differences observed in the example programs. This involves using the identified arguments effectively and employing appropriate programming constructs to create a stack of objects. Here's a detailed description of the structure of the meta-program:

### 1. Function Definition:
The function should be defined with the arguments identified earlier, including the main ones (label, num_objects, initial_position, displacement_y, half_size) and optional ones (random_offset).

### 2. List Initialization:
Initialize an empty list `objs` to store the created objects, ensuring that the created objects can be accessed and manipulated as needed.

### 3. Loop to Create Objects:
Use a loop structure to iterate through the number of objects (`num_objects`). This loop will handle the creation and placement of each object in the stack.

### 4. Object Creation:
Within the loop, create the object using the `create(label, half_size)` function. Each object will have the same type and dimensions as specified by the arguments.

### 5. Position Calculation:
Calculate the position for each object. This involves using the initial position (`initial_position`) and applying the vertical displacement incrementally.

- **Base Position**: Start with the initial position for the first object.
- **Incremental Displacement**: For each subsequent object, add the vertical displacement (`displacement_y`) to the y-coordinate of the previous object's position.
- **Optional Random Offset**: Apply a random offset to the vertical displacement if the `random_offset` argument is provided, simulating a less perfect stack.

### 6. Moving Objects:
Use the `move()` function to place each object at the calculated position. Ensure that the x and z coordinates remain constant for the stack along the y-axis.

### 7. Append to List:
Append each created and moved object to the `objs` list.

### 8. Return the List:
After the loop completes, return the `objs` list containing all the objects in the stack.

### Programming Constructs:
- **Loop (for loop)**: To iterate through the number of objects.
- **Arithmetic Operations**: For calculating the positions.
- **Conditional Handling**: To apply optional random offsets.
- **List Operations**: For storing and returning the objects.

### Pseudo-Code:
Below is a pseudo-code representation to help visualize the structure:

```python
def create_stack(label, num_objects, initial_position, displacement_y, half_size, random_offset=None):
    objs = []
    initial_x, initial_y, initial_z = initial_position

    for i in range(num_objects):
        # Calculate vertical position with optional random offset
        y_position = initial_y + i * displacement_y
        if random_offset:
            y_position += random.uniform(-random_offset, random_offset)

        # Create and move the object
        obj = create(label, half_size)
        move(obj, initial_x, y_position, initial_z)

        # Append to the list
        objs.append(obj)

    return objs
```

### Summary:
The meta-program is structured to use the provided arguments to create a stack of objects. It leverages a loop for object creation and placement, arithmetic operations for position calculations, optional randomness for variability, and list operations to manage the created objects. This structure ensures the generality and flexibility needed to recreate the given example spatial motifs and generate new variations within the "stack" motif.
\end{minted}

\mypara{Meta-program generation.}

\noindent Asking LLM to generate a \metaprog for the motif type \prompt{stack} based on the reasoning it just did using the prompt \prompt{generalize\_low\_level}. The LLM is asked to also use docstring and include comments.

\begin{minted}[breaklines, breakafter=d, fontsize=\scriptsize]{text}
Take a deep breath and follow the steps below carefully.
You have identified the arguments and the structure of the meta-program.
Now, write the meta-program that can create various spatial motifs of "<MOTIF_TYPE>".
Here is a meta-program of the same motif type you created in the past:
```python
<PAST_META_PROGRAM>
```
You can use this as a reference to write the new meta-program, but you can also write it from scratch.
Refer to the arguments you identified and the structure you described to write the meta-program.
Remember that the meta-program should be able to recreate the spatial motifs of the example programs as closely as possible.
Avoid taking plain lists of object attributes, such as coordinates, sizes, and rotations, as arguments, as this defeats the purpose of using a meta-program to capture fundamental spatial patterns.
Include a docstring at the beginning of the meta-program to describe the purpose of the function and the arguments it takes.
Use comments to explain your code step by step.
Please respond with code only.
\end{minted}

\noindent Response from LLM (python)
\begin{minted}[breaklines, breakafter=d, fontsize=\scriptsize]{python}
import random

def create_stack(label, num_objects, initial_position, displacement_y, half_size, random_offset=None):
    """
    Create a stack of objects with the specified parameters.

    Args:
    label (str): Type of object to create (e.g., 'plate').
    num_objects (int): Number of objects in the stack.
    initial_position (list[float]): Initial position [x, y, z] of the first object.
    displacement_y (float): Vertical displacement between objects.
    half_size (list[float]): Dimensions of each object as [half_width, half_height, half_depth].
    random_offset (float, optional): Range for random variation in vertical displacement.

    Returns:
    list: A list of created objects.
    """
    objs = []
    initial_x, initial_y, initial_z = initial_position

    for i in range(num_objects):
        # Calculate vertical position with optional random offset
        y_position = initial_y + i * displacement_y
        if random_offset is not None:
            y_position += random.uniform(-random_offset, random_offset)

        # Create and move the object
        obj = create(label, half_size)
        move(obj, initial_x, y_position, initial_z)

        # Append the object to the list
        objs.append(obj)

    return objs
\end{minted}

\mypara{Meta-program validation.}

\noindent Asking LLM to create function calls to the meta-program it just created for each of the motif programs, using the prompt \prompt{generalize\_low\_level\_batch\_recreate}. 
\begin{minted}[breaklines, breakafter=d, fontsize=\scriptsize]{text}
Using the meta-program you wrote, what are the function calls needed to recreate each of the example programs you analyzed previously?
Write one function call for each example program, that recreates the spatial motif of the program as closely as possible.
Ignore the programs you hypothesized and only write function calls for the example programs I provided.
Response with a json-like text structure with the example program order integers as keys and the function calls as values.
For example, {"1": "function_call(program1_args)", "2": "function_call(program2_args)", ...}
\end{minted}

\noindent Response from LLM (json)
\begin{minted}[breaklines, breakafter=d, fontsize=\scriptsize]{json}
{
  "1": "create_stack('plate', 7, [0.0, 0.0, 0.0], -0.00757, [0.08909, 0.0143, 0.08853])"
}
\end{minted}

\noindent All validation checks were passed after iteration 1. If the validation did not pass, the LLM is asked to try to correct for any errors using the \prompt{generalize\_low\_level\_feedback} prompt.

\mypara{Meta-program documentation.}
\noindent Asking LLM to refine the comments in the meta-program using the prompt \prompt{generalize\_refine\_comments}
\begin{minted}[breaklines, breakafter=d, fontsize=\scriptsize]{text}
You have successfully written the meta-program to generalize the spatial motif of "<MOTIF_TYPE>".
The latest function calls you provided are able to recreate the spatial motifs of the example programs.
Please update the docstring and comments in the meta-program to provide as much information as possible about the expected arguments.
If the program contains well-defined parts that can be explained, include comments to describe these parts - how they contribute to the spatial motif and how they differ from other parts.
At inference time, the meta-program will be the sole source of information for recreating this spatial motif.
Therefore, these comments will be crucial for understanding how the meta-program should be used.
Include at least one example function call in the comments to show how the meta-program should be called.
Please respond with code only.
\end{minted}

\noindent Response from LLM (python)
\begin{minted}[breaklines, breakafter=d, fontsize=\scriptsize]{python}
import random

def create_stack(label, num_objects, initial_position, displacement_y, half_size, random_offset=None):
    """
    Create a stack of objects with the specified parameters.

    Args:
    label (str): Type of object to create (e.g., 'plate').
    num_objects (int): Number of objects in the stack.
    initial_position (list[float]): Initial position [x, y, z] of the first object.
    displacement_y (float): Vertical displacement between objects.
    half_size (list[float]): Dimensions of each object as [half_width, half_height, half_depth].
    random_offset (float, optional): Range for random variation in vertical displacement. Default is None.

    Returns:
    list: A list of created objects.

    Example Call:
    objs = create_stack('plate', 7, [0.0, 0.0, 0.0], -0.00757, [0.08909, 0.0143, 0.08853])

    This example recreates a stack of seven plates with the given initial position, displacement, and dimensions.
    """

    # Initialize an empty list to store the created objects
    objs = []

    # Extract initial position components for readability
    initial_x, initial_y, initial_z = initial_position

    # Iterate over the number of objects to be created in the stack
    for i in range(num_objects):
        # Calculate vertical position with optional random offset
        y_position = initial_y + i * displacement_y
        if random_offset is not None:
            y_position += random.uniform(-random_offset, random_offset)

        # Create the object with the specified label and dimensions
        obj = create(label, half_size)

        # Move the object to its calculated position
        move(obj, initial_x, y_position, initial_z)

        # Append the created and moved object to the list
        objs.append(obj)

    # Return the list of created objects
    return objs
\end{minted}

\noindent The meta-program provided by the LLM is then stored in the meta-program library with the motif type of \prompt{stack}.